\documentclass[margin,12pt]{article}
\usepackage[catalan,english]{babel}
\usepackage{lmodern}
\usepackage[T1]{fontenc}
\usepackage{latexsym}
\usepackage{slashbox}
\usepackage{amsfonts}
\usepackage{amsmath}
\usepackage{amssymb}
\usepackage[mathscr]{eucal}
\usepackage{a4wide}
\usepackage{graphicx}
\usepackage[usenames]{color}
\usepackage{xcolor}
\usepackage{bbold}
\usepackage{bbm}
\usepackage{authblk}
\numberwithin{equation}{section}
\usepackage{multirow}
\RequirePackage{hyperref}
\hypersetup{
	linktocpage,
    colorlinks,
    citecolor=blue,
    filecolor=black,
    linkcolor=blue,
    urlcolor=blue,
    }

\def\be{\begin{equation}}
\def\ee{\end{equation}}
\def\bea{\begin{eqnarray}}
\def\eea{\end{eqnarray}}
\def\ben{\begin{enumerate}}
\def\een{\end{enumerate}}

\title{\vspace{-1.25cm}
\bf{A dispersive analysis of the pion vector form factor and $\tau^{-}\to K^{-}K_{S}\nu_{\tau}$ decay}}
\author[a,b]{Sergi Gonz\`{a}lez-Sol\'{i}s\thanks{sgonzalez@itp.ac.cn}}
\author[c]{Pablo Roig\thanks{proig@fis.cinvestav.mx}}
\affil[a]{{\it{CAS Key Laboratory of Theoretical Physics, Institute of Theoretical Physics, Chinese Academy of Sciences, Beijing 100190, China}}}
\affil[b]{{\it{Department of Physics, Indiana University, Bloomington, IN 47405, USA\newline
Center for Exploration of Energy and Matter, Indiana University, Bloomington, IN 47408, USA}}}
\affil[c]{{\it{Departamento de F\'{i}sica, Centro de Investigaci\'{o}n y Estudios Avanzados del IPN\\ Apdo. Postal 14-740, 07000 Ciudad de M\'{e}xico, M\'{e}xico}}}

\begin{document}\maketitle

\begin{abstract}
We explore the potential of a combined analysis of the decays $\tau^{-}\to\pi^{-}\pi^{0}\nu_{\tau}$ and $\tau^{-}\to K^{-}K_{S}\nu_{\tau}$ in the determination of the $\rho(1450)$ and $\rho(1700)$ resonance properties
in the frame of resonance Chiral Theory supplemented by dispersion relations.
On the one hand, we take advantage of the very precise data on the modulus squared of the pion vector form factor $|F_{V}^{\pi}|^{2}$ obtained by Belle to carry out a very dedicated analysis of the region where these resonances come up into play.
Our study provides an improved treatment of the systematic theoretical errors and, as a most important result, we conclude that they dominate over the fit uncertainties in the determination of the $\rho(1450)$ and $\rho(1700)$ pole parameters and tend to be larger than in other determinations quoted in the literature where these errors were ignored or underestimated.
The results of our analysis are summarized in numerical tables for the form factor modulus and phase, including both statistical and systematic errors, that can be found as ancillary material of this paper.
As a byproduct, we also determine the low-energy observables of the pion vector form factor and the $\rho$-pole position.
On the other hand, we benefit from the recent experimental data for the transition $\tau^{-}\to K^{-}K_{S}\nu_{\tau}$ released by BaBar to perform a first analysis of its decay spectrum and discuss the role of these resonances in this decay.
We point out that higher-quality data on the $K^{-}K_{S}$ decay channel will allow compete with the $|F_{V}^{\pi}|^{2}$ ones and improve the determination of the $\rho(1450)$ and $\rho(1700)$ resonance parameters as a result of a combined analysis.
We hope our study to be of interest for present and future experimental analysis of these decays.

\end{abstract}

\section{Introduction}\label{section1}

Tau lepton decays into a neutrino and hadrons provide a privileged scenario to investigate the non-perturbative regime of QCD under rather clean conditions since half of the transition is purely electroweak and can be computed straightforwardly.
Such advantageous framework is used to improve our understanding of the hadronization of QCD currents as well as for determining the physical parameters, mass and width, of the intermediate resonances produced in the decay \cite{Pich:2013lsa}.
The strong dynamics is encoded within the hadronic matrix element which, in turn, is given in terms of form factors.
As it is well-known, Chiral Perturbation Theory (ChPT) \cite{Gasser:1983yg,Gasser:1984ux} provides a successful description of these form factors valid at very low energies $E<<M_{\rho}$, where $M_{\rho}$ is the $\rho(770)$ resonance mass.
However, as one approaches the resonance region $E\sim M_{\rho}$, ChPT ceases to provide a good description of the physics and the resonance fields shall be explicitly incorporated into the description as new degrees of freedom.
This is the aim of Resonance Chiral Theory (RChT) \cite{Ecker:1988te}, an effective field theory with resonance fields built in.
RChT has been used by different groups as the initial setup approach to describe two meson tau decays providing a good description of the experimental measurements after being supplemented by arguments of analyticity and unitarity through dispersion relations.
For example, the analyses of the $\pi^{-}\pi^{0}$ \cite{Guerrero:1997ku,Pich:2001pj,Dumm:2013zh,Celis:2013xja} and $K\pi$ \cite{Jamin:2006tk,Jamin:2008qg,Boito:2008fq,Boito:2010me} decay channels were found to be in a nice agreement with the rich experimental data provided by experiments.

One of the purposes of this work is to extend our series of dedicated analyses of two meson tau decays based on the framework of RChT supplemented by dispersion relations i.e. $\tau^{-}\to\pi^{-}\pi^{0}\nu_{\tau}$ \cite{Dumm:2013zh}, $\tau^{-}\to K_{S}\pi^{-}\nu_{\tau}$ and $\tau^{-}\to K^{-}\eta^{(\prime)}\nu_{\tau}$ \cite{Escribano:2014joa,Escribano:2013bca}, and $\tau^{-}\to\pi^{-}\eta^{(\prime)}\nu_{\tau}$ \cite{Escribano:2016ntp}, to the $K^{-}K_{S}$ final state meson system.
The topic is of timely interest due to the recent measurement of the $\tau^{-}\to K^{-}K_{S}\nu_{\tau}$ decay spectrum released by the BaBar collaboration \cite{BaBar:2018qry}.
This measurement is based on a sample of 223741 events and significantly improves the mass spectrum measured by CLEO in 1996 \cite{Coan:1996iu} were only 100 events in the $\tau^{-}\to K^{-}K_{S}\nu_{\tau}$ were selected.
The threshold for the decay $\tau\to K^{-}K_{S}\nu_{\tau}$ opens around 1000 MeV which is $\sim$100 MeV larger than $M_{\rho}+\Gamma_{\rho}$, a characteristic energy scale for the $\rho(770)$-dominance region.
This implies that the $K^{-}K_{S}$ decay mode is not sensitive to the $\rho(770)$ peak, and consequently not useful to study its properties, but rather enhances its sensitivity to the properties of the heavier copies $\rho(1450)$ and $\rho(1700)$.


Also, a precise theoretical determination of the two-pion vector form factor within the SM (with a robust error band) is needed to increase the accuracy of the search for non-standard interactions in semileptonic weak charged currents \cite{Garces:2017jpz,Miranda:2018cpf,Cirigliano:2018dyk}. 
This information is included as supplementary material of this paper (see Appendix \ref{Supplementary material}).

These facts motivate the present work where we intend to demonstrate that a reanalysis of the $\pi^{-}\pi^{0}$ and $K^{-}K_{S}$ decay spectra at present B-factories such Belle-II \cite{Kou:2018nap} could help improve notably the knowledge of the $\rho(1450)$ and $\rho(1700)$ resonance properties.

First, we reexamine the pion vector form factor focusing our effort on the improvement of the description of the energy region where the $\rho(1450)$ and $\rho(1700)$ come up into play. 
Our analysis is based on a three-times-subtracted dispersion relation.
For the required input of the form factor phase entering the dispersive integral, we rely on Watson's theorem \cite{Watson:1954uc}\footnote{Watson's theorem applied to the pion vector form factor tell us that the form factor phase equals that of the two-pion scattering within the elastic region.} and take advantage of the well-known parametrization of the $\pi\pi$ scattering phase shift existent in the literature, to drive the form factor phase up to 1 GeV.
Above 1 GeV, we get a model for the phase from the exponential Omn\`{e}s representation that we explicate in detail in section \ref{section2}.
This parametrization establishes the framework of our form factor description.
However, we will also consider a number of variants to this approach that will be used to assess the (important) role of the systematic uncertainties in the extraction of the $\rho(1450)$ and $\rho(1700)$ resonance parameters that have been usually ignored or underestimated in the literature so far.
Altogether results, to the best of our knowledge, in the most dedicated analysis of the intermediate- and high-energy region of the pion vector form factor experimental data to date.
Also, the impact on the low-energy observables of the pion vector form factor is addressed and discussed as a byproduct of our approach.
  
Second, we built a parametrization for the kaon vector form factor in a similar fashion as for the pion one and perform a first analysis of the $\tau^{-}\to K^{-}K_{S}\nu_{\tau}$ BaBar experimental measurement.
The role of each participating resonance is discussed and the corresponding parameters are extracted from fits to data.

Finally, in view of the findings obtained from our analyses of the individual $\pi^{-}\pi^{0}$ and $K^{-}K_{S}$ channels, we perform a combined analysis to both data sets to see what can be learned.
From our study, we can anticipate that although the $\tau^{-}\to K^{-}K_{S}\nu_{\tau}$ BaBar data supersede the old CLEO data, still the precision is not sufficiently good enough to compete with the pion vector form factor modulus squared data from Belle. 
In all, we hope our study to be of interest for present and future experimental analysis of these decays.

This article is structured as follows.
In section \ref{section2} we provide a bottom-up review of the pion vector form factor organized according to the fulfillment of unitarity and analyticity constraints.
For illustrative purposes, we start with the ChPT calculation at order $\mathcal{O}{(p^{4})}$ and follow by the explicit inclusion of vector resonance states.
For our study, we will consider three resonances i.e. $\rho(770),\rho(1450)$ and the $\rho(1700)$, and then submit the form factor to a unitarization procedure through the Omn\`{e}s integral that leads to the Omn\`{e}s exponentiation of the form factor. 
This parametrization allows us to get a model for the phase of the form factor valid up to the mass of the $\tau$.
This phase is then inserted into a three-times-subtracted dispersion relation that completes our representation of the form factor, and the corresponding model parameters are fitted to the Belle measurement of the modulus squared of the pion vector form factor $|F_{V}^{\pi}(s)|^{2}$.
Predictions and fits to the BaBar $\tau^{-}\to K^{-}K_{S}\nu_{\tau}$ decay spectrum measurement are discussed in section \ref{section3}.
In section \ref{section4}, we perform joint fits to the decays $\tau^{-}\to\pi^{-}\pi^{0}\nu_{\tau}$ and $\tau^{-}\to K^{-}K_{S}\nu_{\tau}$ and 
finally, our conclusions are presented in section \ref{conclusions}. 

\section{The pion vector form factor}\label{section2}

The pion vector form factor has been measured in $e^{+}e^{-}\to\pi^{+}\pi^{-}$ \cite{Amendolia:1986wj,Aloisio:2004bu,Akhmetshin:2006bx,Aubert:2009ad,Ambrosino:2010bv,Ablikim:2015orh,Anastasi:2017eio} and in $\tau\to\pi^{-}\pi^{0}\nu_{\tau}$ \cite{Anderson:1999ui,Fujikawa:2008ma} and widely studied in the literature \cite{Guerrero:1997ku,Pich:2001pj,Dumm:2013zh,Celis:2013xja,Bijnens:1998fm,Bijnens:2002hp,Oller:2000ug,Bruch:2004py,Hanhart:2012wi,SanzCillero:2002bs,Pich:2010sm,Colangelo:2018mtw} since it enters the description of many physical processes.
As it is the main object concerning our analysis, we will thus provide in the following a brief, but detailed review of its calculation following a bottom-up approach according to the fulfillment of unitarity and analyticity constraints.
We will start with the Chiral Perturbation Theory calculation at $\mathcal{O}(p^{4})$ to follow with the explicit inclusion of resonances as degrees of freedom.
The $\pi\pi$ and $KK$ final-state interactions will be then resummed to all order through an Omn\`{e}s exponentiation. 
Finally, a three-times-subtracted dispersion relation completes our form factor representation.

\subsection{Exponential Omn\`{e}s representation}\label{exponentialrepresentation}

The pion vector form factor $F_{V}^{\pi}(s)$ is defined through the hadronic matrix element of the vector current between the vacuum and the pion pair
\begin{equation}
\langle\pi^{0}\pi^{-}|\bar{d}\gamma^{\mu}u|0\rangle=\sqrt{2}\left(p_{\pi^{-}}-p_{\pi^{0}}\right)^{\mu}F_{V}^{\pi}(s)\,,
\label{HadronicMatrixElementPion}
\end{equation}
where $s=(p_{\pi^{-}}+p_{\pi^{0}})^{2}$.

At very low-energies, the pion vector form factor is well described by ChPT.
The resulting calculation at $\mathcal{O}(p^{4})$ reads \cite{Gasser:1984ux} (in the limit of exact isospin symmetry)

\begin{equation}
F_{V}^{\pi}(s)|_{\rm{ChPT}}=1+\frac{2L_{9}^{r}(\mu)}{F_{\pi}^{2}}s-\frac{s}{96\pi^{2}F_{\pi}^{2}}\left[A_{\pi}(s,\mu^{2})+\frac{1}{2}A_{K}(s,\mu^{2})\right]\,,
\label{FFChPT}
\end{equation}
where $L_{9}^{r}(\mu)$ is one of the renormalized low-energy couplings constants in the $\mathcal{O}(p^{4})$ chiral Lagrangian, and $A_{P}(s,\mu^{2})$ are two-pseudoscalars loop-functions (whose renormalization-scale dependence cancels out with the one in $L_{9}^{r}$) accounting for the unitary corrections given by 

\begin{equation}
A_{P}(s,\mu^{2})=\log\frac{m_{P}^{2}}{\mu^{2}}+\frac{8m_{P}^{2}}{s}-\frac{5}{3}+\sigma_{P}^{3}(s)\log\left(\frac{\sigma_{P}(s)+1}{\sigma_{P}(s)-1}\right)\,,
\end{equation}
where
\begin{equation}
\sigma_{P}(s)=\sqrt{1-\frac{4m_{P}^{2}}{s}}\,.
\end{equation}

The validity of ChPT is restricted to very low energies, and as one approaches the region where the influence of new degrees of freedom, the lightest meson resonances, becomes important, ChPT ceases to provide a good description.
Resonance Chiral Theory partly cures this limitation incorporating such resonances explicitly. 
For the case concerning us, the $\rho(770)$ resonance dominates the form factor.
At leading order in powers of $1/N_{C}$, which it is $\mathcal{O}(p^{4})$ in the chiral expansion, the result is given by \cite{Ecker:1988te}
\begin{equation}
F_{V}^{\pi}(s)=1+\frac{F_{V}G_{V}}{F_{\pi}^{2}}\frac{s}{M_{\rho}^{2}-s}\,,
\end{equation}
where $F_{V}$ and $G_{V}$ measure, respectively, the strength of the $\rho V^{\mu}$ and $\rho\pi\pi$ couplings, with $V^{\mu}$ being the quark vector current.
Assuming that the form factor vanishes when $s\to\infty$ one gets the condition
\begin{equation}
F_{V}G_{V}=F_{\pi}^{2}\,,
\end{equation}
that yields the usual Vector Meson Dominance (VMD) in the zero-width approximation
\begin{equation}
F_{V}^{\pi}(s)=\frac{M_{\rho}^{2}}{M_{\rho}^{2}-s}\,.
\label{VMD}
\end{equation}
Re-expanding Eq.\,(\ref{VMD}) in $s$ and comparing with the polynomial part of its ChPT counterpart in Eq.\,(\ref{FFChPT}) one gets an estimate for the $\mathcal{O}(p^{4})$ chiral coupling $L_{9}^{r}$
\begin{equation}
L_{9}^{r}(M_{\rho})=\frac{F_{V}G_{V}}{2M_{\rho}^{2}}=\frac{F_{\pi}^{2}}{2M_{\rho}^{2}}\backsimeq7.2\cdot10^{-3}\,,
\end{equation}
which is in very good agreement with the value extracted from phenomenology.
This result shows explicitly that the $\rho(770)$ contribution is indeed the dominant physical effect in the pion vector form factor.

An improved realization of the pion vector form factors stems from combining Eqs.\,(\ref{FFChPT}) and (\ref{VMD}), which yields:
\begin{equation}
F_{V}^{\pi}(s)=\frac{M_{\rho}^{2}}{M_{\rho}^{2}-s}-\frac{s}{96\pi^{2}F_{\pi}^{2}}\left[A_{\pi}(s,\mu^{2})+\frac{1}{2}A_{K}(s,\mu^{2})\right]\,,
\end{equation}
where the first term (VMD) is the dominant one in $1/N_{C}$ and resums an infinite number of local contributions in ChPT to all orders, while the second term includes the loop contributions that are next order in $1/N_{C}$.

In the spirit of Refs.\,\cite{Guerrero:1997ku,Jamin:2006tk,Jamin:2008qg}, one can do better and perform a resummation of the $\pi\pi$ and $KK$ final-state interactions to all orders relying on unitarity and analyticity constraints.
This leads to the Omn\`{e}s exponentiation of the full loop function  
\begin{equation}
F_{V}^{\pi}(s)=\frac{M_{\rho}^{2}}{M_{\rho}^{2}-s}\exp\Big\lbrace-\frac{s}{96\pi^{2}F_{\pi}^{2}}\left[A_{\pi}(s,\mu^{2})+\frac{1}{2}A_{K}(s,\mu^{2})\right]\Big\rbrace\,.
\label{FFChPTExp}
\end{equation}
However, the previous expression still has an obvious defect, it cannot describe the energy region of the peak of the $\rho$-meson.
For that, it is necessary to incorporate its width.
The energy dependent width of the $\rho$ is related to the imaginary part of the loop function and is given by \cite{GomezDumm:2000fz}
\begin{eqnarray}
\Gamma_{\rho}(s)&=&-\frac{M_{\rho}s}{96\pi^{2}F_{\pi}^{2}}{\rm{Im}}\left[A_{\pi}(s)+\frac{1}{2}A_{K}(s)\right]\nonumber\\[1ex]
&=&\frac{M_{\rho}s}{96\pi F_{\pi}^{2}}\left[\sigma_{\pi}^{3}(s)\theta(s-4m_{\pi}^{2})+\frac{1}{2}\sigma_{K}^{3}(s)\theta(s-4m_{K}^{2})\right]\,.
\label{RhoWidth}
\end{eqnarray}
In order to account for the $\rho$-resonance width, we insert $\Gamma_{\rho}(s)$ in the propagator of the $\rho$ in Eq.\,(\ref{FFChPTExp}) arriving at:
\begin{equation}
F_{V}^{\pi}(s)=\frac{M_{\rho}^{2}}{M_{\rho}^{2}-s-iM_{\rho}\Gamma_{\rho}(s)}\exp\Big\lbrace-\frac{s}{96\pi^{2}F_{\pi}^{2}}{\rm{Re}}\left[A_{\pi}(s,\mu^{2})+\frac{1}{2}A_{K}(s,\mu^{2})\right]\Big\rbrace\,,
\label{ExponentialParam}
\end{equation}
where, in order to avoid double counting of the imaginary part of the loop functions, only the real part of the loops is kept in the exponential.
We would like to point out here that strict analyticity and unitarity is only maintained if the real part of the loop integral function is resummed in the propagator together with the imaginary part \cite{Boito:2008fq}. 
Up to $\mathcal{O}(p^{4})$ in the chiral expansion, resumming the real part in the propagator or in the exponential is fully equivalent and differences between the two approaches start to appear at $\mathcal{O}(p^{6})$.
This effect, however, is seen to be numerically negligible in the $\pi\pi$ and $K\pi$ systems.

In all, equation (\ref{ExponentialParam}) provides a suitable description of the $\rho$-dominance region.  
However, the precise measurement of the $\tau^{-}\to\pi^{-}\pi^{0}\nu_{\tau}$ decay spectrum by the Belle collaboration \cite{Fujikawa:2008ma} indicates that heavier resonance excitations i.e. $\rho^{\prime}\equiv\rho(1450)\,,\rho^{\prime\prime}\equiv\rho(1700)$, contribute and cannot be simply neglected.
We incorporate these higher excited resonances into the description in an analogous fashion as we have done for the $\rho$.
The resulting expression takes the form:
\begin{eqnarray}
F_{V}^{\pi}(s)&=&\frac{M_{\rho}^{2}+s\left(\gamma e^{i\phi_{1}}+\delta e^{i\phi_{2}}\right)}{M_{\rho}^{2}-s-iM_{\rho}\Gamma_{\rho}(s)}\exp\Bigg\lbrace {\rm{Re}}\Bigg[-\frac{s}{96\pi^{2}F_{\pi}^{2}}\left(A_{\pi}(s)+\frac{1}{2}A_{K}(s)\right)\Bigg]\Bigg\rbrace\nonumber\\[2mm]
&&-\gamma\frac{s\,e^{i\phi_{1}}}{M_{\rho^{\prime}}^{2}-s-iM_{\rho^{\prime}}\Gamma_{\rho^{\prime}}(s)}\exp\Bigg\lbrace-\frac{s\Gamma_{\rho^{\prime}}(M_{\rho^{\prime}}^{2})}{\pi M_{\rho^{\prime}}^{3}\sigma_{\pi}^{3}(M_{\rho^{\prime}}^{2})}{\rm{Re}}A_{\pi}(s)\Bigg\rbrace\nonumber\\[2mm]
&&-\delta\frac{s\,e^{i\phi_{2}}}{M_{\rho^{\prime\prime}}^{2}-s-iM_{\rho^{\prime\prime}}\Gamma_{\rho^{\prime\prime}}(s)}\exp\Bigg\lbrace-\frac{s\Gamma_{\rho^{\prime\prime}}(M_{\rho^{\prime\prime}}^{2})}{\pi M_{\rho^{\prime\prime}}^{3}\sigma_{\pi}^{3}(M_{\rho^{\prime\prime}}^{2})}{\rm{Re}}A_{\pi}(s)\Bigg\rbrace\,,
\label{FFExpThreeRes}
\end{eqnarray}
where the coefficients $\gamma$ and $\delta$ measure the relative weight between the contributions of the different resonances while the phases $\phi_{1}$ and $\phi_{2}$ stand for the corresponding interference, and with
\begin{eqnarray}
\Gamma_{\rho^{\prime},\rho^{\prime\prime}}(s)&=&\Gamma_{\rho^{\prime},\rho^{\prime\prime}}\frac{s}{M_{\rho^{\prime},\rho^{\prime\prime}}^{2}}\frac{\sigma^{3}_{\pi}(s)}{\sigma^{3}_{\pi}(M_{\rho^{\prime},\rho^{\prime\prime}}^{2})}\theta(s-4m_{\pi}^{2})\,.
\end{eqnarray}
Notice that, for the energy-dependent width of the $\rho^{\prime}$ and $\rho^{\prime\prime}$ given in the previous equation, we have assumed that these resonances only decay into $\pi\pi$ since in our resonance chiral framework there is no warranty that other intermediate states contribute in the proportion given in Eq.\,(\ref{RhoWidth}) for the $\rho$-meson width.
However, one can still explore a model including other intermediate states in a similar fashion and see what can be learned.
We shall come back to this in section \ref{dispersiverepresentation}.

At this point, we would like to anticipate that the exponential Omn\`{e}s representation of the pion vector form factor given by Eq.\,(\ref{FFExpThreeRes}) will be used as input for the parametrization of the form factor phase entering the dispersive approach described in section \ref{dispersiverepresentation} which, in turn, will be used to get the central results of this work.
In particular, the extracted phase will be employed to describe the energy region from 1 GeV to $m_{\tau}$, and matched smoothly to the $\pi\pi$ scattering phase-shift solution of the Roy equations of \cite{GarciaMartin:2011cn} at 1 GeV. 
By doing this matching, sensitivity is lost to whether or not the real part of the loop function is resummed into the propagator denominator or kept into the exponential as discussed few lines above since the differing numerical results are tiny.

In the following, however, we would like first to prove this parametrization against experimental data as a warm-up.
In total, we have nine unknown parameters, $\lbrace M_{\rho},\gamma,\phi_{1},M_{\rho^{\prime}},\\\Gamma_{\rho^{\prime}},\delta,\phi_{2},M_{\rho^{\prime\prime}},\Gamma_{\rho^{\prime\prime}}\rbrace$, that can be inferred from fits to the measured modulus squared of the pion vector form factor extracted from the Belle $\tau^{-}\to\pi^{-}\pi^{0}\nu_{\tau}$ measurement \cite{Fujikawa:2008ma}. 
The resulting fit parameters are\footnote{In all our fits throughout the paper, whenever the pion  form factor is involved, we employ $F_{\pi}=92.316$ MeV, which is the central value using the restriction $\sqrt{2}|V_{ud}|F_{\pi}=(127.13\pm0.02\pm0.13)$ MeV from the 2018 PDG edition \cite{PhysRevD.98.030001} with the 2006 PDG reported value $|V_{ud}|=0.97377\pm0.00027$ used by Belle.} 
\begin{eqnarray}
\label{FitpionFFExp}
& &M_{\rho}\,=\,775.2(4)\,\rm{MeV}\,,\quad \gamma\,=\,0.15(4)\,,\quad \phi_{1}\,=\,-0.36(24)\,,\nonumber\\[2mm]
& & M_{\rho^{\prime}}\,=\,1438(39)\,\rm{MeV}\,,\quad \Gamma_{\rho^{\prime}}\,=\,535(63)\,\rm{MeV}\,,\quad \delta\,=-0.12(4)\,,\quad\phi_{2}\,,=-0.02(45)\,,\nonumber\\[2mm]
& & M_{\rho^{\prime\prime}}\,=\,1754(91)\,\rm{MeV}\,,\quad \Gamma_{\rho^{\prime\prime}}\,=\,412(102)\,\rm{MeV}\,,
\end{eqnarray}
with a $\chi^{2}$/d.o.f $=48.9/53\sim0.92$, and where the associated uncertainty is the statistical error resulting from the fit.
Notice, however, that $M_{\rho,\rho^{\prime},\rho^{\prime\prime}}$ and $\Gamma_{\rho^{\prime},\rho^{\prime\prime}}$ are model input parameters and do not correspond to the physical resonance mass and width.
In order to extract the physical resonance pole parameters one should compute the pole position in the complex $s_{R}$ plane according to $\sqrt{s_{R}}=M_{R}-\frac{i}{2}\Gamma_{R}$.
By doing so, the pole parameters associated to the $\rho$, $\rho^{\prime}$ and $\rho^{\prime\prime}$  resonances are then found to be
\begin{eqnarray}
\label{PoleParam}
& & M^{\rm{pole}}_{\rho}\,=\,762.0(3)\,\rm{MeV}\,,\quad \Gamma^{\rm{pole}}_{\rho}\,=\,143.0(2)\,\rm{MeV}\,,\nonumber\\[2mm]
& & M^{\rm{pole}}_{\rho^{\prime}}\,=\,1366(38)\,\rm{MeV}\,,\quad \Gamma^{\rm{pole}}_{\rho^{\prime}}\,=\,488(48)\,\rm{MeV}\,,\\[2mm]
& & M^{\rm{pole}}_{\rho^{\prime\prime}}\,=\,1718(82)\,\rm{MeV}\,,\quad \Gamma^{\rm{pole}}_{\rho^{\prime\prime}}\,=\,397(88)\,\rm{MeV}\,,\nonumber
\end{eqnarray}
where the uncertainties are calculated by assuming a Gaussian error propagation while simultaneously varying the corresponding unphysical masses and widths in Eq.\,(\ref{FitpionFFExp}).

\begin{figure}[h!]
\begin{center}
\includegraphics[scale=0.85]{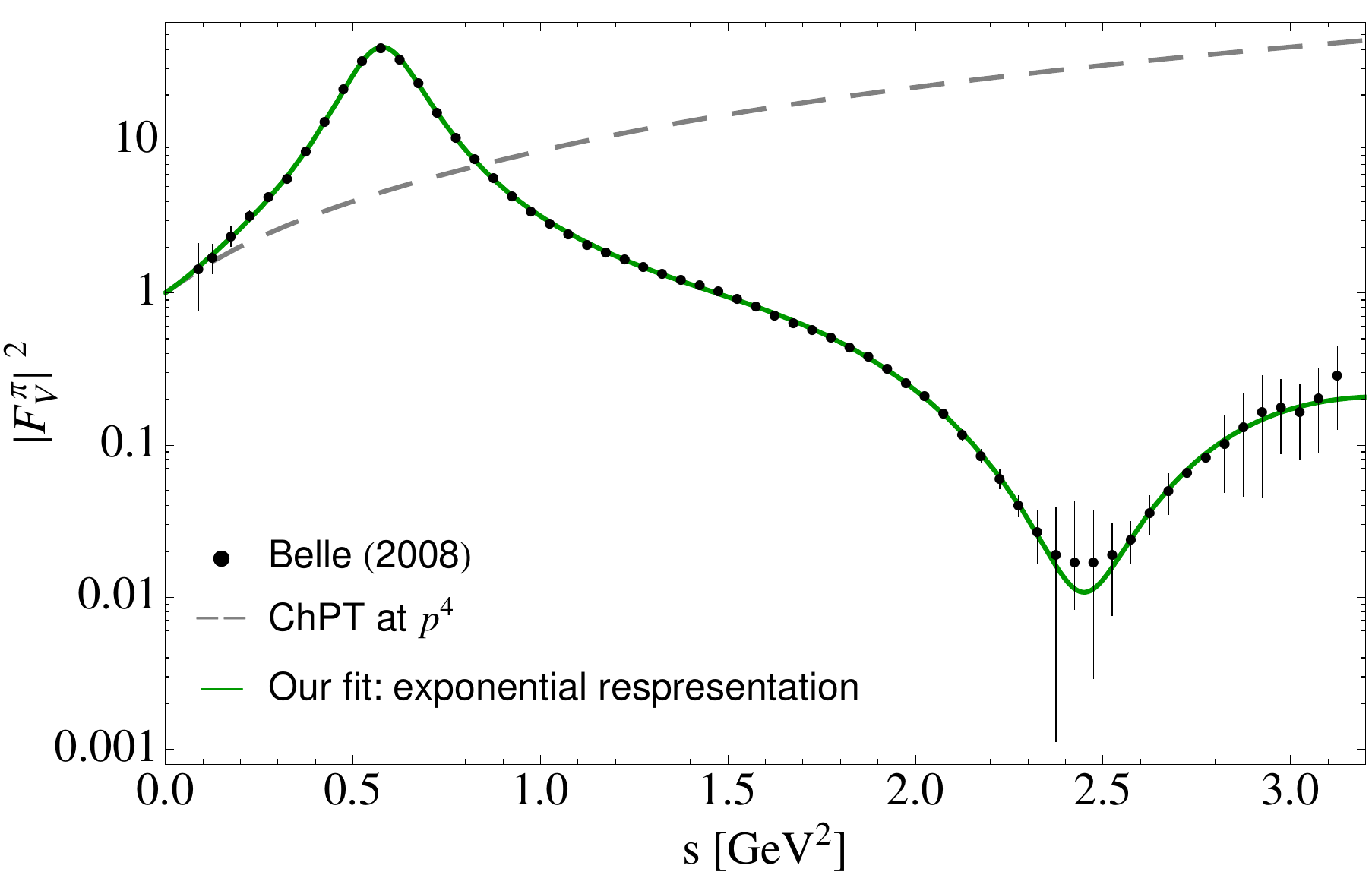}
\caption{\label{FitExp}Belle measurement of the modulus squared of the pion vector form factor $|F_{V}^{\pi}|^{2}$ (black solid circles) \cite{Fujikawa:2008ma} as compared to our fit results (solid green line) as presented in Eq.\,(\ref{FitpionFFExp}). The ChPT calculation at $\mathcal{O}(p^{4})$ is also shown for illustration (dashed gray line).}
\end{center}
\end{figure}

The resulting form factor corresponding to our fit is displayed in Fig.\,\ref{FitExp} (solid green line) confronted to Belle data \cite{Fujikawa:2008ma}.
As can be seen from the plot and the $\chi^{2}$/d.o.f, the agreement with data is very satisfactory. 
In the figure, the ChPT calculation at $\mathcal{O}(p^{4})$ (cf.\,Eq.\,(\ref{FFChPT})) is also shown (dashed gray line) for illustrative purposes\footnote{For simplicity, in the figure we do not represent the ChPT calculation at $\mathcal{O}(p^{6})$ \cite{Bijnens:1998fm,Bijnens:2002hp} since including higher-order chiral corrections increases very little the energy region where the data is described well.}. 

\subsection{Dispersive representation}\label{dispersiverepresentation}

A Cauchy dispersion relation representation of the pion vector form factor is fully determined by the discontinuity across the cut along the positive real axis.
Contributions to the discontinuity arise every time an intermediate state production threshold opens starting at $s_{\rm{th}}=4m_{\pi}^{2}$, the lightest possible contribution.
The elastic approximation is confined to the two-pion contribution to the discontinuity and neglects heavier intermediate state contributions.
In this limit, Watson's theorem \cite{Watson:1954uc} states that the phase of the form factor equals that of the elastic $\pi\pi$ scattering phase and the form factor admits an analytic solution given in terms of the phase shift.
Such a solution is the well-known Omn\`{e}s equation \cite{Omnes:1958hv} that, in terms of the $I=1$ $P$-wave $\pi\pi$ scattering phase shift $\delta_{1}^{1}(s)$ concerning us, reads
\begin{equation}
F_{V}^{\pi}(s)=\Omega(s)=\exp{\left(\frac{s}{\pi}\int_{4m_{\pi}^{2}}^{\infty} ds^{\prime}\frac{\delta_{1}^{1}(s^{\prime})}{s^{\prime}\left(s^{\prime}-s\right)}\right)}\,.
\label{OmnesEq}
\end{equation}
The phase $\delta_{1}^{1}(s)$ entering the dispersive integral encodes the physics of the $\rho$-meson and it is known with an excellent precision in the elastic region $s\le1$ GeV$^{2}$ from the solutions of Roy equations \cite{GarciaMartin:2011cn,Caprini:2011ky} that are valid roughly up to $s_{0}=1.3$ GeV.
However, uncertainties associated to the input of $\delta_{1}^{1}(s)$ can be estimated between the region of 1.3 GeV and the mass of the $\tau$ lepton $\sim1.8$ GeV.
The precise Belle measurement of the $\tau\to\pi^{-}\pi^{0}\nu_{\tau}$ decay distribution \cite{Fujikawa:2008ma} indicates relevant contributions from the excited resonances $\rho^{\prime}$ and $\rho^{\prime\prime}$ that cannot be simply neglected.
To include them, we adopt the form factor representation given in Eq.\,(\ref{FFExpThreeRes}) to get a model for the form factor phase including the three participating resonances.
This phase can be extracted from the relation
\begin{equation}
\tan\delta_{1}^{1}(s)=\frac{{\rm{Im}}F^{\pi}_{V}(s)}{{\rm{Re}}F^{\pi}_{V}(s)}\,,
\label{PhaseModel}
\end{equation}
that is only valid in the $\tau$ decay region $(s<m_{\tau}^{2})$.
For the high-energy region $(s>m_{\tau}^{2})$ we guide smoothly the phase to $\pi$ \cite{Moussallam:1999aq} to ensure the correct asymptotic $1/s$ fall-off of the form factor \cite{Lepage:1979zb}.

The phase that we will use for our analysis consists in matching smoothly at 1 GeV the phase as extracted in Eq.\,(\ref{PhaseModel}) to the phase-shift solution of the Roy equations of Ref.\,\cite{GarciaMartin:2011cn}\footnote{Another successful parametrization of the phase shift is given in Ref.\,\cite{Caprini:2011ky}. 
We only consider one of these parametrizations \cite{GarciaMartin:2011cn} since both agree rather well.}.
This procedure is inspired by, but slightly different than, the ones followed in Ref.\,\cite{Schneider:2012ez} where a fit to data is first performed with the exponential Omn\`{e}s representation of the pion vector form factor without kaon loops (cf.\,Eq.\,(\ref{FFExpThreeRes})) and then matched to the phase-shift solution of the Roy equations.
For our fits, we follow the representation of the pion vector form factor outlined in Refs.\,\cite{Dumm:2013zh,Celis:2013xja} and write a thrice-subtracted dispersion relation 
\begin{equation}
F_{V}^{\pi}(s)=\exp\left[\alpha_{1}s+\frac{\alpha_{2}}{2}s^{2}+\frac{s^{3}}{\pi}\int_{4m_{\pi}^{2}}^{s_{\rm{cut}}}ds^{\prime}\frac{\delta_{1}^{1}(s^{\prime})}{(s^{\prime})^{3}(s^{\prime}-s-i0)}\right]\,,
\label{FFthreesub}
\end{equation}
where $\alpha_{1}$ and $\alpha_{2}$ are two subtraction constants that can be related to chiral low-energy observables, namely the squared charged pion radius $\langle r^{2}\rangle_{V}^{\pi}$ and the coefficients of $\mathcal{O}(s^{2})$ and $\mathcal{O}(s^{3})$ terms in the chiral expansion, $c_{V}^{\pi}$ and $d_{V}^{\pi}$, respectively, appearing in the low-energy expansion of the form factor:
\begin{equation}
F_{V}^{\pi}(s)=1+\frac{1}{6}\langle r^{2}\rangle_{V}^{\pi}s+c_{V}^{\pi}s^{2}+d_{V}^{\pi}s^{3}+\cdots\,.
\label{FFexpansion}
\end{equation}
Explicitly, the relations for the linear and quadratic slope parameters $\langle r^{2}\rangle_{V}^{\pi}$ and $c_{V}^{\pi}$ read
\begin{equation}
\langle r^{2}\rangle_{V}^{\pi}=6\alpha_{1}\,,\quad c_{V}^{\pi}=\frac{1}{2}\left(\alpha_{2}+\alpha_{1}^{2}\right)\,.
\label{FFexpansion2}
\end{equation}
These subtraction constants can be calculated theoretically through the sum rule
\begin{equation}
\alpha_{k}=\frac{k!}{\pi}\int_{4m_{\pi}^{2}}^{s_{\rm{cut}}}ds^{\prime}\frac{\delta_{1}^{1}(s^{\prime})}{s^{\prime k+1}}\,.
\label{SumRule}
\end{equation}
For our analysis, however, we treat them as free parameters to be determined from fits to data. This has the advantage that they turn out to be less model dependent.
Higher slope parameters can be computed from the previous sum rule. 
For example, the cubic slope parameters $d_{V}^{\pi}$ can be obtained through 
\begin{equation}
d_{V}^{\pi}=\frac{1}{6}\left(\alpha_{3}+3\alpha_{1}\alpha_{2}+\alpha_{1}^{3}\right)\,.
\end{equation}

The use of a three-times dispersion relation in Eq.\,(\ref{FFthreesub}) makes the fit less sensitive to the higher-energy region of the dispersive integral where the phase is less well-known.
In total, we have ten free parameters\footnote{The parametrization for the phase shift $\delta_{1}^{1}(s)$ of Ref.\,\cite{GarciaMartin:2011cn} contains a parameter for the $\rho$-meson mass, that we name $m_{\rho}$, that denotes the energy at which the phase shift passes through $\pi/2$ (and therefore it shall not be confused with real part of the pole of the $\rho$) and its quoted value is $m_{\rho}=773.6(9)$ MeV.
For our study, in a first approximation we fix the model input parameter for the $\rho$-meson mass, $M_{\rho}$ in Eq.\,(\ref{FFExpThreeRes}), to this value.
However, will also test the sensitivity of our fits to this parameter by allowing it to float.} entering $F_{V}^{\pi}(s)$ to be determined by a fit to the Belle data.
Regarding the integral cutoff $s_{\rm{cut}}$, one should take a value as large as possible so as not to spoil the {\it{a priori}} infinite interval of integration and to avoid the effects of the spurious singularities generated due to the upper limit being finite, but low enough that the phase is well known within the interval.
The parameters resulting from the fits are given in Table \ref{FitBelle} as Fit 1 for four representative values of $s_{\rm{cut}}$, namely $m_{\tau}^{2}$ (third column), $4$ GeV$^{2}$ (fourth column), $10$ GeV$^{2}$ (fifth column) and finally the $s_{\rm{cut}}\to\infty$ limit (last column). 
The choice of $s_{\rm{cut}}=m_{\tau}^{2}$ is motivated by the fact that the model used to get the phase, Eq.\,(\ref{PhaseModel}), is only valid within the $\tau$ decay region $s\leq s_{\rm{cut}}\sim m_{\tau}^{2}$ and beyond that point the dispersive integral has no physical content.
The resulting form factor corresponding to this cutoff generates, as mentioned above, a singularity at $s=m_{\tau}^{2}$ after bending the form factor shape in the preceding neighborhood region.
As a consequence, the high-energy data points are not described well and, in turn, the values for the resonance parameters should be taken with great care.
In fact, this fit is seen very sensitive to such singularities and the resulting parameters are found to be strongly correlated with unstable associated uncertainties.
Therefore, we consider the results with $s_{\rm{cut}}=m_{\tau}^{2}$ only for illustrative purposes throughout the paper.
Our reference fit corresponds to $s_{\rm{cut}}=4$ GeV$^{2}$ (fourth column in Table \ref{FitBelle}) since this value of the cutoff deals well with the imbalance mentioned above.
This does not mean that we know the phase shift up to that point of the integration interval but rather that the chosen cutoff is large enough that avoids, to large extent, the effects caused by the spurious singularity that arises at $s=s_{\rm{cut}}$.
The results obtained by varying $s_{\rm{cut}}$ in Table \ref{FitBelle} will be used to assess the systematic uncertainties of our fit results obtained with $s_{\rm{cut}}=4$ GeV$^{2}$ \footnote{We would like to note the slightly low $\chi^{2}$/d.o.f. that in general, and in line with Ref.\,\cite{Celis:2013xja}, we find along the fits of this section. This may indicate that there are too many free parameters to fit eventually, but as each of them has a physical meaning, it is reasonable to keep them all.}.
We shall return to a discussion on the integral cutoff below.

\begin{figure}[h!]
\begin{center}
\includegraphics[scale=0.85]{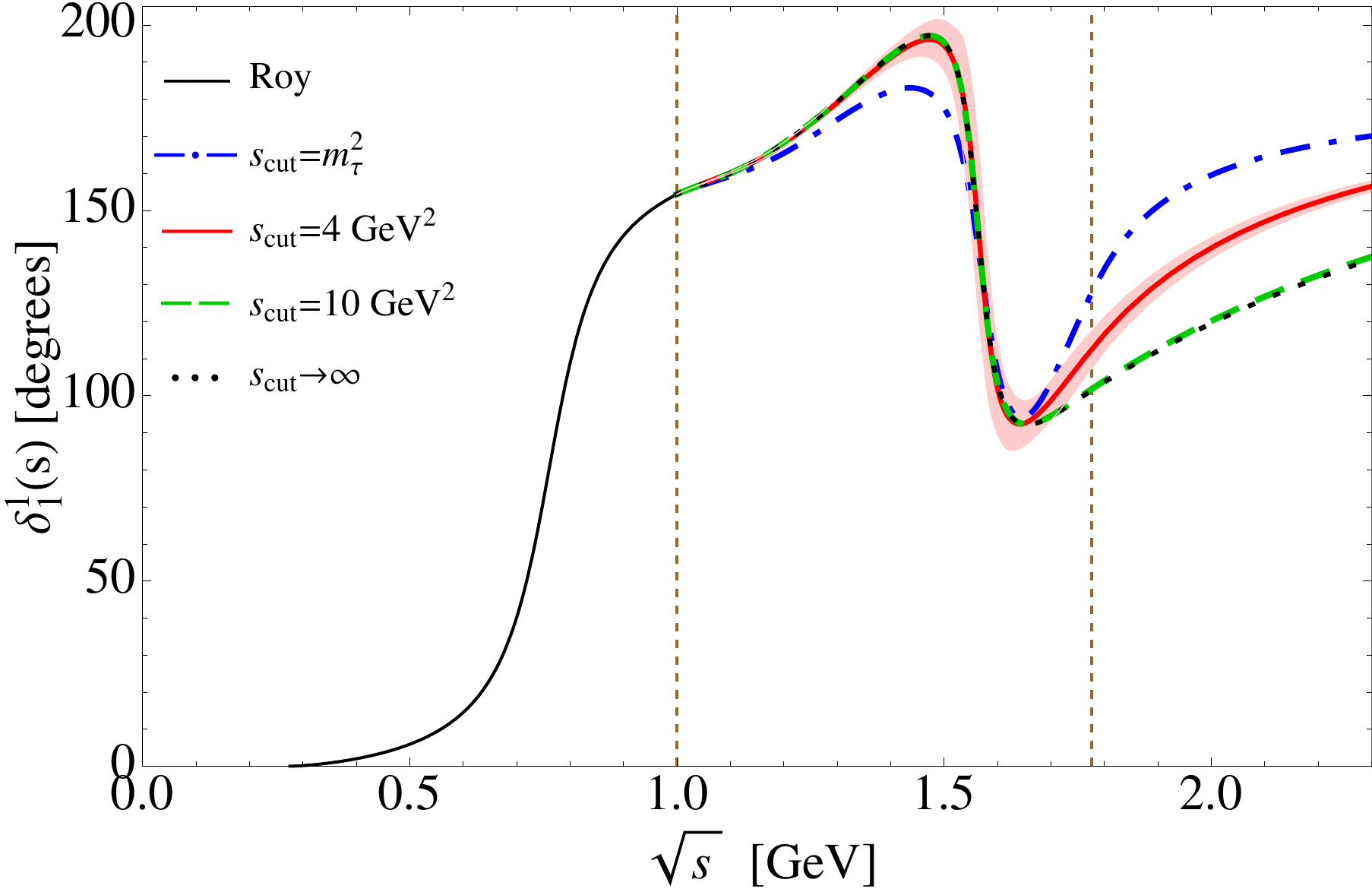}
\caption{\label{PhaseShift}Form factor phase as extracted from fits to the $|F_{V}^{\pi}|^{2}$ Belle data \cite{Fujikawa:2008ma} according to the representation presented in Eq.\,(\ref{FFthreesub}) for four representative values of $s_{\rm{cut}}$ in the dispersive integral. 
The $P$-wave $\pi\pi$ phase shift solution of the Roy equations \cite{GarciaMartin:2011cn} is used up to 1 GeV (solid black curve) and the variations due to $s_{\rm{cut}}$ are given by the dot-dashed blue $(s_{\rm{cut}}=m_{\tau}^{2}$), solid red $(s_{\rm{cut}}=4$ GeV$^{2})$, dashed green $(s_{\rm{cut}}=10$ GeV$^{2})$ and dotted black $(s_{\rm{cut}}\to\infty)$ curves, respectively. 
The two vertical dashed brown lines are placed at 1 GeV and $\sqrt{s}=m_{\tau}$, and denote, respectively, the range where the phase shift from $\pi\pi$ scattering is used and the validity of the parametrization of the form factor phase shift.
All phases are smoothly guided to $\pi$ for $s>m_{\tau}$.
See main text for details.}
\end{center}
\end{figure}

\begin{figure}[h!]
\begin{center}
\includegraphics[scale=0.85]{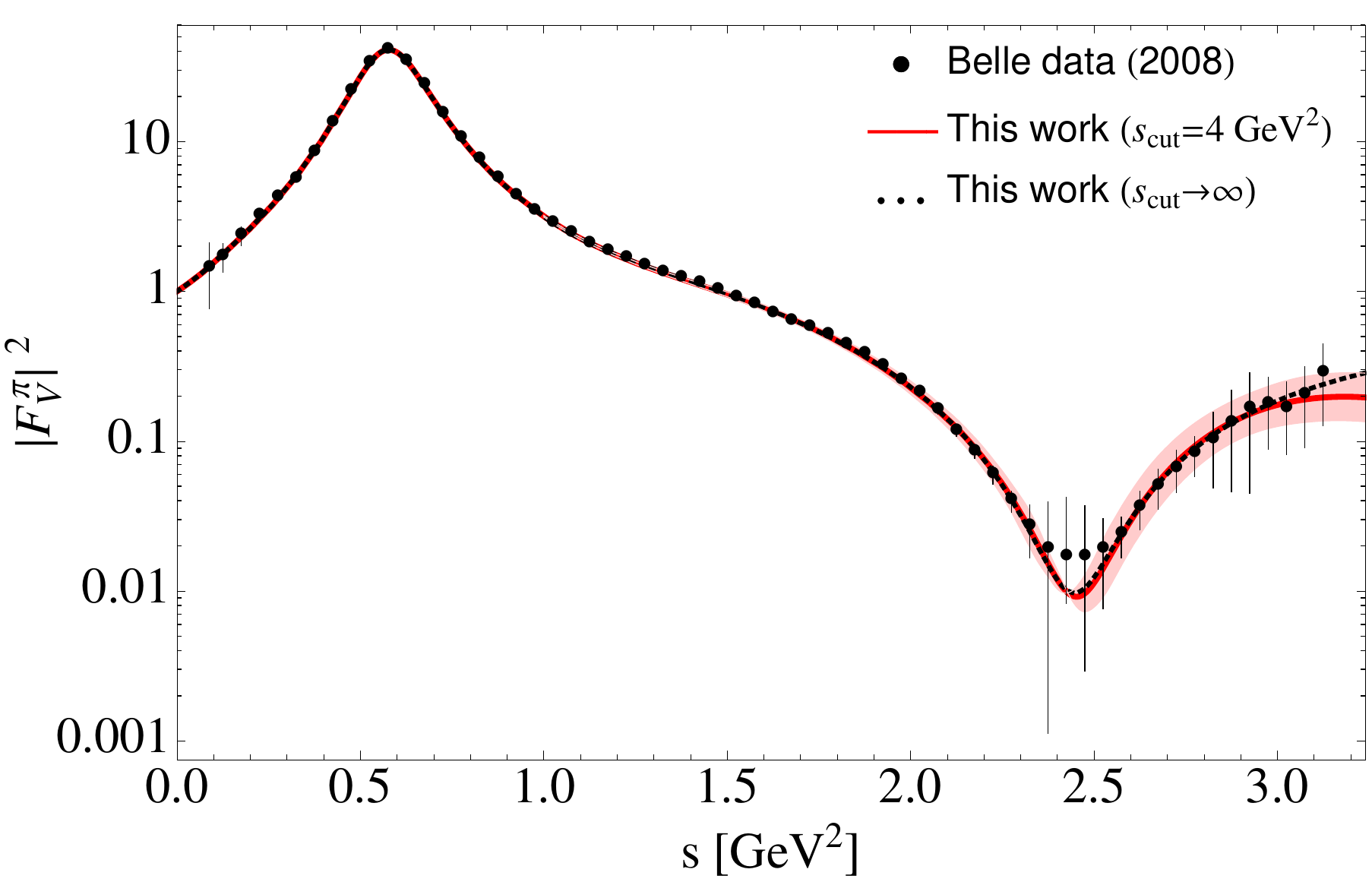}
\caption{\label{FitDispersive}Belle measurement of the absolute value squared of the pion vector form factor $|F_{V}^{\pi}|^{2}$ (black filled circles) \cite{Fujikawa:2008ma} as compared to our fit results as presented in Table\,\ref{FitBelle} for $s_{\rm{cut}}=4$ GeV$^{2}$ (red solid line) and for $s_{\rm{cut}}\to\infty$ (black dotted curve).}
\end{center}
\end{figure}

In Fig.\,\ref{PhaseShift}, we show the resulting phase shifts for the chosen $s_{\rm{cut}}$. 
The phase shift solution of the Roy equations is given by the solid black curve while the variations due to $s_{\rm{cut}}$ are given by the dot-dashed blue $(s_{\rm{cut}}=m_{\tau}^{2}$), solid red $(s_{\rm{cut}}=4$ GeV$^{2})$, dashed green $(s_{\rm{cut}}=10$ GeV$^{2})$ and dotted black $(s_{\rm{cut}}\to\infty)$ curves, respectively.
In the figure, the statistical uncertainty associated to our reference fit $(s_{\rm{cut}}=4$ GeV$^{2})$ is also shown by the light red error band.
Also, there are two brown dashed vertical lines shown in the figure.
They are placed at 1 GeV and $\sqrt{s}=m_{\tau}$ and denote, respectively, the range where the phase shift from $\pi\pi$ scattering is used and the validity of the form factor phase shift parametrization as extracted through Eq.\,(\ref{PhaseModel}).
In Fig.\,\ref{FitDispersive}, we provide a graphical account of the resulting form factor for $s_{\rm{cut}}=4$ GeV$^{2}$ (solid red curve and light red error band) and $s_{\rm{cut}}\to\infty$ (dotted black curve) compared to Belle data \cite{Fujikawa:2008ma} . 
As can be seen from the figure and the corresponding $\chi^{2}$/d.o.f, excellent agreement with experimental data is seen with all data points.

In order to optimize the phase shift in the fit to the pion form factor measurement from $\tau^{-}\to\pi^{-}\pi^{0}\nu_{\tau}$ data, we have also run fits allowing to float the parameter for the $\rho$-meson mass entering $\delta_{1}^{1}(s)$. 
The corresponding fit results are contained in Table \ref{FitBelle} as Fit 1-$\rho$ and the resulting parameters are found to be in general accordance with those of Fit 1.

\begin{table}[h!]
\begin{center}
  \begin{tabular}{|l|l|l|l|l|c|c|c|c|c|c|c|c|c|c|c|c|c|}
\hline
    &\multirow{2}{*}{Parameter} &  
    \multicolumn{4}{c|}{\multirow{1}{*}{{$s_{\rm{cut}}$ [GeV$^{2}$]}}} \\ 
    \cline{3-6}
     Fits&&$m_{\tau}^{2}$ & $4$ (reference fit)& $10$ & $\infty$\\ \cline{1-6}
Fit 1&$\alpha_{1}$ [GeV$^{-2}$]&$1.87(1)$&$1.88(1)$&$1.89(1)$&$1.89(1)$\cr
&$\alpha_{2}$  [GeV$^{-4}$]&$4.40(1)$&$4.34(1)$&$4.32(1)$&$4.32(1)$\cr
 &$m_{\rho}$ [MeV]&$=773.6(9)$&$=773.6(9)$&$=773.6(9)$&$=773.6(9)$\cr
 &$M_{\rho}$ [MeV]&$=m_{\rho}$&$=m_{\rho}$&$=m_{\rho}$&$=m_{\rho}$\cr
&$M_{\rho^{\prime}}$ [MeV]&$1365(15)$&$1376(6)$&$1313(15)$&$1311(5)$\cr
&$\Gamma_{\rho^{\prime}}$ [MeV]&$562(55)$&$603(22)$&$700(6)$&$701(28)$\cr
&$M_{\rho^{\prime\prime}}$[MeV] &$1727(12)$&$1718(4)$&$1660(9)$&$1658(1)$\cr
&$\Gamma_{\rho^{\prime\prime}}$ [MeV]&$278(1)$&$465(9)$&$601(39)$&$602(3)$\cr
&$\gamma$ &$0.12(2)$&$0.15(1)$&$0.16(1)$&$0.16(1)$\cr
&$\phi_{1}$ &$-0.69(1)$&$-0.66(1)$&$-1.36(10)$&$-1.39(1)$\cr
&$\delta$ &$-0.09(1)$&$-0.13(1)$&$-0.16(1)$&$-0.17(1)$\cr
&$\phi_{2}$ &$-0.17(5)$&$-0.44(3)$&$-1.01(5)$&$-1.03(2)$\cr
\cline{2-6}
&$\chi^{2}$/d.o.f&$1.47$&$0.70$&$0.64$&$0.64$\cr
  \hline     
  Fit 1-$\rho$&$\alpha_{1}$ [GeV$^{-2}$]&$1.88(1)$&$1.88(1)$&$1.89(1)$&$1.88(1)$\cr
&$\alpha_{2}$ [GeV$^{-4}$]&$4.37(3)$&$4.34(1)$&$4.31(3)$&$4.34(1)$\cr
&$m_{\rho}$ [MeV]&$773.9(3)$&$773.8(3)$&$773.9(3)$&$773.9(3)$\cr
 &$M_{\rho}$ [MeV]&$=m_{\rho}$&$=m_{\rho}$&$=m_{\rho}$&$=m_{\rho}$\cr
&$M_{\rho^{\prime}}$ [MeV]&$1382(71)$&$1375(11)$&$1316(9)$&$1312(8)$\cr
&$\Gamma_{\rho^{\prime}}$ [MeV]&$516(165)$&$608(35)$&$728(92)$&$726(26)$\cr
&$M_{\rho^{\prime\prime}}$[MeV] &$1723(1)$&$1715(22)$&$1655(1)$&$1656(8)$\cr
&$\Gamma_{\rho^{\prime\prime}}$ [MeV]&$315(271)$&$455(16)$&$569(160)$&$571(13)$\cr
&$\gamma$ &$0.12(13)$&$0.16(1)$&$0.18(2)$&$0.17(1)$\cr
&$\phi_{1}$ &$-0.56(35)$&$-0.69(1)$&$-1.40(19)$&$-1.41(8)$\cr
&$\delta$ &$-0.09(3)$&$-0.13(1)$&$-0.17(4)$&$-0.17(3)$\cr
&$\phi_{2}$ &$-0.19(69)$&$-0.45(12)$&$-1.06(10)$&$-1.05(11)$\cr
\cline{2-6}
&$\chi^{2}$/d.o.f&$1.09$&$0.70$&$0.63$&$0.66$\cr
  \hline     
  \end{tabular}
\caption{\label{FitBelle}\small{Results for the fits obtained with a three-times-subtracted dispersion relation including three vector resonances in $F_{V}^{\pi}(s)$ according to Eq.\,(\ref{FFthreesub}) for four representative values of $s_{\rm{cut}}$ in the dispersive integral.}}
\end{center}
\end{table}

For our reference fit, we have also probed the theoretical uncertainty associated to the choice of the matching point with the phase-shift solution of the Roy equations.
The dependence of the fitted parameters on the matching point 
is explored through the fits that we collect in Table \ref{FitBelleMatching} as Fit I 
for a fixed 
$\rho$ mass parameter.
The matching point values $0.85$ GeV (third column), $0.90$ GeV (fourth column), $0.95$ GeV (fifth column) and 1 GeV (last column) are used (the matching point $1$ GeV corresponds to our reference fit in the fourth column of Table \ref{FitBelle} and is repeated here for ease of comparison).
A look at the results of this table reveals that, in general, all the parameters are rather stable against variations of the matching point, although the $\rho^{\prime\prime}$ width is sensitive to these variations with a tendency of becoming larger as it decreases.

\begin{table}[h!]
\begin{center}
  \begin{tabular}{|l|l|l|l|l|c|c|c|c|c|c|c|c|c|c|c|c|c|}
\hline
    &\multirow{2}{*}{Parameter} &  
    \multicolumn{4}{c|}{\multirow{1}{*}{{Matching point [GeV]}}} \\ 
    \cline{3-6}
     Fits&&$0.85$ & $0.9$& $0.95$ & $1$ (reference fit)\\ \cline{1-6}
Fit I&$\alpha_{1}$ [GeV$^{-2}$]&$1.88(1)$&$1.88(1)$&$1.88(1)$&$1.88(1)$\cr
&$\alpha_{2}$ [GeV$^{-4}$]&$4.35(1)$&$4.35(1)$&$4.34(1)$&$4.34(1)$\cr
 &$m_{\rho}$ [MeV]&$=773.6(9)$&$=773.6(9)$&$=773.6(9)$&$=773.6(9)$\cr
 &$M_{\rho}$ [MeV]&$=m_{\rho}$&$=m_{\rho}$&$=m_{\rho}$&$=m_{\rho}$\cr
&$M_{\rho^{\prime}}$ [MeV]&$1394(6)$&$1374(8)$&$1351(5)$&$1376(6)$\cr
&$\Gamma_{\rho^{\prime}}$ [MeV]&$592(19)$&$583(27)$&$592(2)$&$603(22)$\cr
&$M_{\rho^{\prime\prime}}$[MeV] &$1733(9)$&$1715(1)$&$1697(3)$&$1718(4)$\cr
&$\Gamma_{\rho^{\prime\prime}}$ [MeV]&$562(3)$&$541(45)$&$486(7)$&$465(9)$\cr
&$\gamma$ &$0.12(1)$&$0.12(1)$&$0.13(1)$&$0.15(1)$\cr
&$\phi_{1}$ &$-0.44(3)$&$-0.60(1)$&$-0.80(1)$&$-0.66(1)$\cr
&$\delta$ &$-0.13(1)$&$-0.13(1)$&$-0.13(1)$&$-0.13(1)$\cr
&$\phi_{2}$ &$-0.38(3)$&$-0.51(2)$&$-0.62(1)$&$-0.44(3)$\cr
\cline{2-6}
&$\chi^{2}$/d.o.f&$0.75$&$0.74$&$0.68$&$0.70$\cr
  \hline     
  \end{tabular}
\caption{\label{FitBelleMatching}\small{Results for the fits obtained with a three-times-subtracted dispersion relation including three vector resonances in $F_{V}^{\pi}(s)$ according to Eq.\,(\ref{FFthreesub}) with $s_{\rm{cut}}=4$ GeV$^{2}$ in the dispersive integral for four representative values of the matching point.}}
\end{center}
\end{table}
We shall now return to the discussion mentioned in section \ref{exponentialrepresentation} about the inclusion of intermediate states other than $\pi\pi$ into the $\rho^{\prime}$ and $\rho^{\prime\prime}$ decay widths.
We next allow such resonances to decay not only in $\pi\pi$ but also into $K\bar{K}$.
In this case, the energy-dependent widths read
\begin{eqnarray}
\Gamma_{\rho^{\prime},\rho^{\prime\prime}}(s)&=&\Gamma_{\rho^{\prime},\rho^{\prime\prime}}\frac{s}{M^{2}_{\rho^{\prime},\rho^{\prime\prime}}}\left[\frac{\sigma^{3}_{\pi}(s)}{\sigma^{3}_{\pi}(M_{\rho^{\prime},\rho^{\prime\prime}}^{2})}\theta(s-4m_{\pi}^{2})+\frac{1}{2}\frac{\sigma^{3}_{K}(s)}{\sigma^{3}_{K}(M_{\rho^{\prime},\rho^{\prime\prime}}^{2})}\theta(s-4m_{K}^{2})\right]\,,
\label{KKintowidths}
\end{eqnarray}
and these are incorporated into the corresponding resonance propagators of Eq.\,(\ref{FFExpThreeRes}), while the real part of the kaon loop is resumed in the exponential in a similar fashion as the pion ones.
The corresponding fit results are gathered in Table \ref{CentralFitDispersiveWidths} as Fit A (second column) and are compared to our reference fit (last column) again repeated here for ease of comparison.
In this case, the $\chi^{2}$ increases by 12 and, as seen, there are two or three parameters that are affected by the fact of including kaons into the decay widths, the rest are seen rather stable upon comparison.
The $\rho^{\prime}$ width decreases by $\sim140$ MeV while the $\rho^{\prime\prime}$ mass increases up to $1775$ MeV.
The $\rho^{\prime\prime}$ width is slightly shifted downwards but the associated error is enlarged.

We have also explored a variant of Eq.\,(\ref{KKintowidths}) that includes, moreover, the contribution $\rho^{\prime}\to\omega\pi$ into the $\rho^{\prime}$ decay width.
In spirit of \cite{Edwards:1999fj}, we write the energy-dependent width as
\begin{eqnarray}
\Gamma_{\rho^{\prime}}(s)=&&\Gamma_{\rho^{\prime}}\frac{s}{M^{2}_{\rho^{\prime}}}\Big[\mathcal{BR}(\rho^{\prime}\to(\pi\pi+KK))\left(\frac{\sigma^{3}_{\pi}(s)}{\sigma^{3}_{\pi}(M_{\rho^{\prime}}^{2})}\theta(s-4m_{\pi}^{2})+\frac{1}{2}\frac{\sigma^{3}_{K}(s)}{\sigma^{3}_{K}(M_{\rho^{\prime}}^{2})}\theta(s-4m_{K}^{2})\right)\nonumber\\[1ex]
&&+{\mathcal{BR}(\rho^{\prime}\to\omega\pi)}\frac{\sigma_{\omega\pi}(s)}{\sigma_{\omega\pi}(M_{\rho^{\prime}}^{2})}\theta(s-(m_{\omega}+m_{\pi})^{2})\Big]\,,
\label{OmegaPiintowidths}
\end{eqnarray}
with
\begin{eqnarray}
\sigma_{\omega\pi}(s)=\frac{1}{s}\sqrt{(s-(m_{\omega}-m_{\pi})^{2})(s-(m_{\omega}+m_{\pi})^{2})}\,,
\end{eqnarray}
and where $\mathcal{BR}(\rho^{\prime}\to(\pi\pi+KK))$ and $\mathcal{BR}(\rho^{\prime}\to\omega\pi)$ are relative branching ratios normalized with their sum equal to one.
We use the PDG estimate $\mathcal{BR}(\rho^{\prime}\to\omega\pi)\equiv\Gamma(\rho^{\prime}\to\omega\pi)/\Gamma^{\rm{total}}_{\rho^{\prime}}\sim0.21$ \cite{PhysRevD.98.030001}, which implies $\mathcal{BR}(\rho^{\prime}\to(\pi\pi+KK))\sim0.79$, and the resulting fit results are presented in Table \ref{CentralFitDispersiveWidths} as Fit B (third column).
In this case, the $\rho^{\prime}$ mass(width) is shifted upwards(downwards) by $65(27)$ MeV with respect to our reference fit, while the $\rho^{\prime\prime}$ width is seen decreased by $116$ MeV.
The other parameters remain rather stable. 
This last exercise serves to have an idea of the potential impact of the channel $\rho^{\prime}\to4\pi$ in $\Gamma_{\rho^{\prime}}(s)$.

\begin{table}[h!]
\begin{center}
  \begin{tabular}{|l|l|l|l|l|c|c|c|c|c|c|c|c|c|c|c|c|c|}
\hline
    \multirow{2}{*}{Parameter} &  
    \multicolumn{3}{c|}{\multirow{1}{*}{{$s_{\rm{cut}}=4$ GeV$^{2}$}}} \\ 
    \cline{2-4}
     & Fit A & Fit B& reference fit\\   \hline    
$\alpha_{1}$ [GeV$^{-2}$]&$1.87(1)$&$1.88(1)$&$1.88(1)$\cr
$\alpha_{2}$  [GeV$^{-4}$]&$4.37(1)$&4.35(1)&$4.34(1)$\cr
 $m_{\rho}$ [MeV]&$=773.6(9)$&$=773.6(9)$&$=773.6(9)$\cr
 $M_{\rho}$ [MeV]&$=m_{\rho}$&$=m_{\rho}$&$=m_{\rho}$\cr
$M_{\rho^{\prime}}$ [MeV]&$1373(5)$&$1441(3)$&$1376(6)$\cr
$\Gamma_{\rho^{\prime}}$ [MeV]&$462(14)$&$576(33)$&$603(22)$\cr
$M_{\rho^{\prime\prime}}$[MeV] &$1775(1)$&$1733(9)$&$1718(4)$\cr
$\Gamma_{\rho^{\prime\prime}}$ [MeV]&$412(27)$&$349(52)$&$465(9)$\cr
$\gamma$ &$0.13(1)$&$0.15(3)$&$0.15(1)$\cr
$\phi_{1}$ &$-0.80(1)$&$-0.53(5)$&$-0.66(1)$\cr
$\delta$ &$-0.14(1)$&$-0.14(1)$&$-0.13(1)$\cr
$\phi_{2}$ &$-0.44(2)$&$-0.46(3)$&$-0.44(3)$\cr
  \hline    
$\chi^{2}$/d.o.f&$0.93$&$0.70$&$0.70$\cr
  \hline     
  \end{tabular}
\caption{\label{CentralFitDispersiveWidths}\small{Results for the fits obtained with a three-times-subtracted dispersion relation including three vector resonances in $F_{V}^{\pi}(s)$ according to Eq.\,(\ref{FFthreesub}) with $s_{\rm{cut}}=4$ GeV$^{2}$ in the dispersive integral with (second column) and without (last column) the $K\bar{K}$ channel in the $\rho^{\prime}$ and $\rho^{\prime\prime}$ energy-dependent resonance widths, and with the additional inclusion of the $\rho^{\prime}\to\omega\pi$ contribution into the $\rho^{\prime}$ width (third column). See main text for details.}}
\end{center}
\end{table}

Finally, we also come back to the discussion on the integral cutoff $s_{\rm{cut}}$ and the corresponding generated singularities.
We found that $s_{\rm{cut}}=4$ GeV$^{2}$ is a suitable value for the integral cutoff in the dispersive integral.
However, we would also like to consider a parametrization that allows both to cut the integral and avoid such singularities.
In the case at hand, we have a parametrization for the form factor phase $\delta_{1}^{1}(s)$ which is valid up to $s=m_{\tau}^{2}$ and the idea is to extend it to the full region in an appropriate way.
We follow Refs.\,\cite{DeTroconiz:2001rip,Yndurain:2002ud} and write
\begin{eqnarray}
F_{V}^{\pi}(s)=f(s)\Sigma(s)\,,
\end{eqnarray}
where the function $f(s)$ is given by a once subtracted dispersion relation, that ensures $f(0)=1$, defined by
\begin{eqnarray}
f(s)=\exp\left[\frac{s}{\pi}\int_{4m_{\pi}^{2}}^{s_{\rm{cut}}}ds^{\prime}\frac{\delta_{1}^{1}(s^{\prime})}{s^{\prime}(s^{\prime}-s)}+\frac{s}{\pi}\int_{s_{\rm{cut}}}^{\infty}ds^{\prime}\frac{\bar{\delta}_{1}^{1}(s^{\prime})}{s^{\prime}(s^{\prime}-s)}\right]\,.
\label{FormFactorCutoff}
\end{eqnarray}
The value of the phase $\bar{\delta}_{1}^{1}(s)$ in Eq.\,(\ref{FormFactorCutoff}) should be such that it avoids the generation of spurious singularities and ensures the $1/s$ behavior of the form factor for $s\to\infty$.
In order to fulfill these properties, we choose a smooth interpolation in $s$ for $\delta_{1}^{1}(s)$ above $s_{\rm{cut}}$ as simply as
\begin{eqnarray}
\bar{\delta}_{1}^{1}(s)=\pi+\left[\delta_{1}^{1}(s_{\rm{cut}})-\pi\right]\frac{s_{\rm{cut}}}{s}\,,
\label{phaseinterpolation}
\end{eqnarray}
so that $\bar{\delta}_{1}^{1}(s_{\rm{cut}})=\delta_{1}^{1}(s_{\rm{cut}})$ and $\delta_{1}^{1}(s)\to\pi$ for large $s$ recovering the $1/s$ fall-off of $F_{V}^{\pi}(s)$.
In this case, the integral going from $s_{\rm{cut}}$ to $\infty$ in Eq.\,(\ref{FormFactorCutoff}) can be calculated explicitly and we arrive at \cite{DeTroconiz:2001rip,Yndurain:2002ud}
\begin{eqnarray}
F_{V}^{\pi}(s)&=&\exp\left[\frac{s}{\pi}\int_{4m_{\pi}^{2}}^{s_{\rm{cut}}}ds^{\prime}\frac{\delta_{1}^{1}(s^{\prime})}{s^{\prime}(s^{\prime}-s)}\right]\nonumber\\[1ex]
&&\times\exp\left[1-\frac{\delta_{1}^{1}(s_{\rm{cut}})}{\pi}\right]\left(1-\frac{s}{s_{\rm{cut}}}\right)^{\left[1-\frac{\delta_{1}^{1}(s_{\rm{cut}})}{\pi}\right]\frac{s_{\rm{cut}}}{s}}\left(1-\frac{s}{s_{\rm{cut}}}\right)^{-1}\Sigma(s)\,.
\label{FormFactorCutoffFinal}
\end{eqnarray}
Regarding the function $\Sigma(s)$, it contains the (inelastic) contributions beyond $s_{\rm{cut}}$ and may be understood as giving the correction to the linear continuation of the phase $\delta_{1}^{1}(s)$ above $s_{\rm{cut}}$ as we did in Eq.\,(\ref{phaseinterpolation}).
It is described by an analytical function on the $s$-plane with a cut from $s_{\rm{cut}}$ to $\infty$ and should be obtained from a model or fitted to experiment since it is largely unknown.
Often, it is parametrized by a conformal transformation that maps the right-hand cut in the complex $s$-plane into the unit circle through
\begin{eqnarray}
\Sigma(s)=\sum_{i=0}^{\infty}a_{i}\omega^{i}(s)\,,
\label{sigma}
\end{eqnarray}
with the variable $\omega(s)$ given by
\begin{eqnarray}
\omega(s)=\frac{\sqrt{s_{\rm{cut}}}-\sqrt{s_{\rm{cut}}-s}}{\sqrt{s_{\rm{cut}}}+\sqrt{s_{\rm{cut}}-s}}\,.
\end{eqnarray}
For our analysis, we take the condition $a_{0}=1$ to ensure $F_{V}^{\pi}(0)=1$.

We next probe the application of Eq.\,(\ref{FormFactorCutoffFinal}) against data for $s_{\rm{cut}}=4$ GeV$^{2}$ with one and two parameters in the expansion of Eq.\,(\ref{sigma}).
The resulting fit parameters are found to be
\begin{eqnarray}
& &a_{1}\,=\,2.99(12)\,,\nonumber\\[2mm]
& & M_{\rho^{\prime}}\,=\,1261(7)\,\rm{MeV}\,,\quad \Gamma_{\rho^{\prime}}\,=\,855(15)\,\rm{MeV}\,,\nonumber\\[2mm]
& & M_{\rho^{\prime\prime}}\,=\,1600(1)\,\rm{MeV}\,,\quad \Gamma_{\rho^{\prime\prime}}\,=\,486(26)\,\rm{MeV}\,,\nonumber\\[2mm]
& &\gamma\,=\,0.25(2)\,,\quad \phi_{1}\,=\,-1.90(6)\,,\nonumber\\[2mm]
& &\delta\,=\,-0.15(1)\,,\quad \phi_{2}\,=\,-1.60(4)\,,
\label{FitBelleCutoff2}
\end{eqnarray}
with a $\chi^{2}$/d.o.f $=32.3/53\sim0.61$ for the one-parameter fit, and 
\begin{eqnarray}
& &a_{1}\,=\,3.03(20)\,,\quad a_{2}=1.04(2.10)\,,\nonumber\\[2mm]
& & M_{\rho^{\prime}}\,=\,1303(19)\,\rm{MeV}\,,\quad \Gamma_{\rho^{\prime}}\,=\,839(102)\,\rm{MeV}\,,\nonumber\\[2mm]
& & M_{\rho^{\prime\prime}}\,=\,1624(1)\,\rm{MeV}\,,\quad \Gamma_{\rho^{\prime\prime}}\,=\,570(99)\,\rm{MeV}\,\nonumber\\[2mm]
& &\gamma\,=\,0.22(10)\,,\quad \phi_{1}\,=\,-1.65(4)\,,\nonumber\\[2mm]
& &\delta\,=\,-0.18(1)\,,\quad \phi_{2}\,=\,-1.34(14)\,,
\label{FitBelleCutoff1}
\end{eqnarray}
with a $\chi^{2}$/d.o.f $=35.6/52\sim0.63$ for the two-parameter fit.
The large uncertainty associated to $a_{2}$ suggests not continuing the expansion to higher orders.

\begin{figure}[h!]
\begin{center}
\includegraphics[scale=0.85]{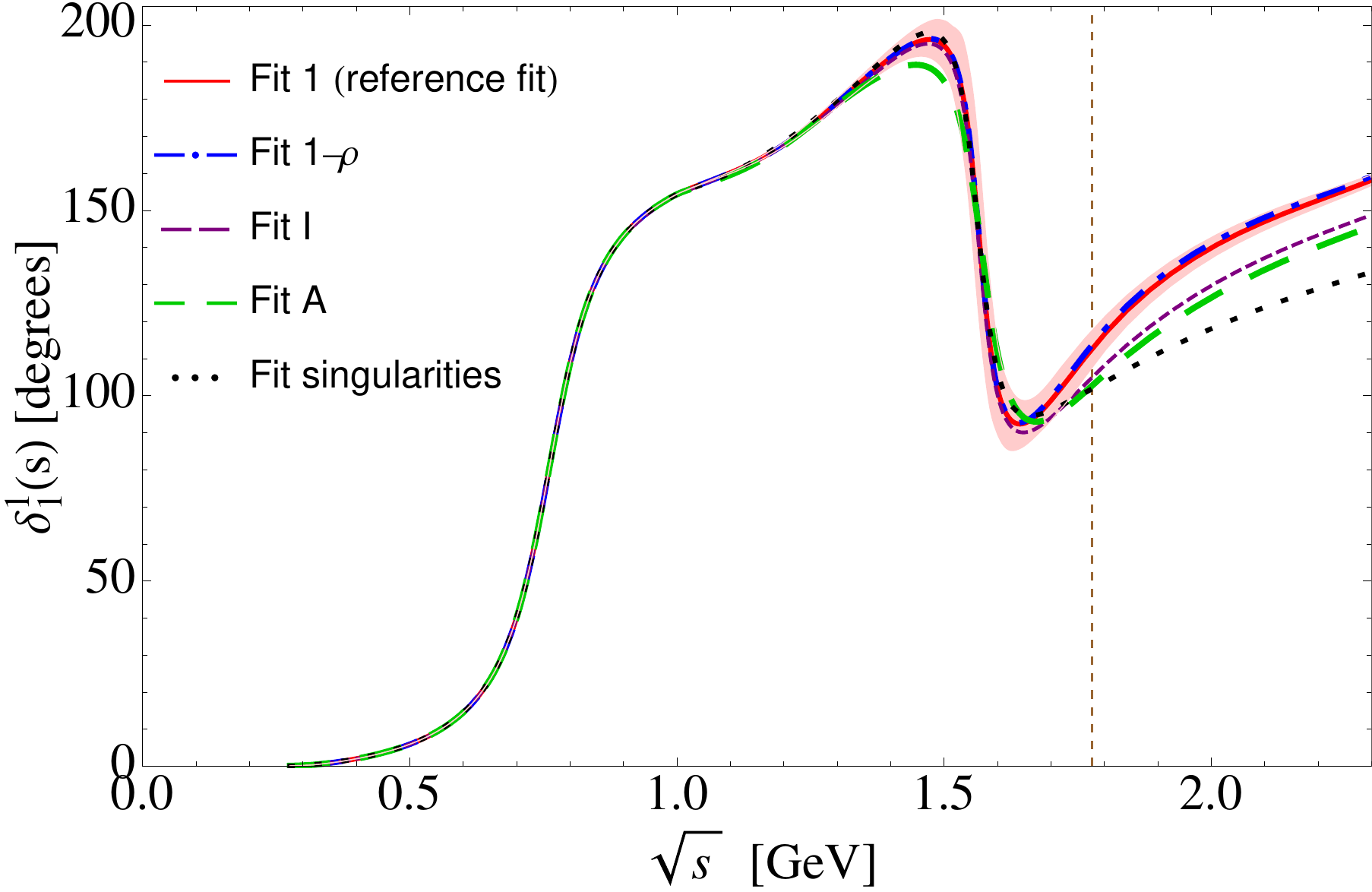}
\caption{\label{FitDispersiveComparisonPhase}Results for the form factor phase shift as extracted from our reference fit in Table\,\ref{FitBelle}, Fit 1 (solid red curve), and from Fit 1-$\rho$ at the matching point of 0.85 GeV (dot-dashed blue curve), Fit I (short dashed purple curve), Fit A (long dashed green curve) and with the fit that avoids singularities (dotted black curve).
The vertical dashed brown line is placed at $m_{\tau}$ and denotes the validity of the parametrization of the form factor phase shift.
All phases are smoothly guided to $\pi$ for $s>m_{\tau}$.
See main text for details.}
\end{center}
\end{figure}

\begin{figure}[h!]
\begin{center}
\includegraphics[scale=0.85]{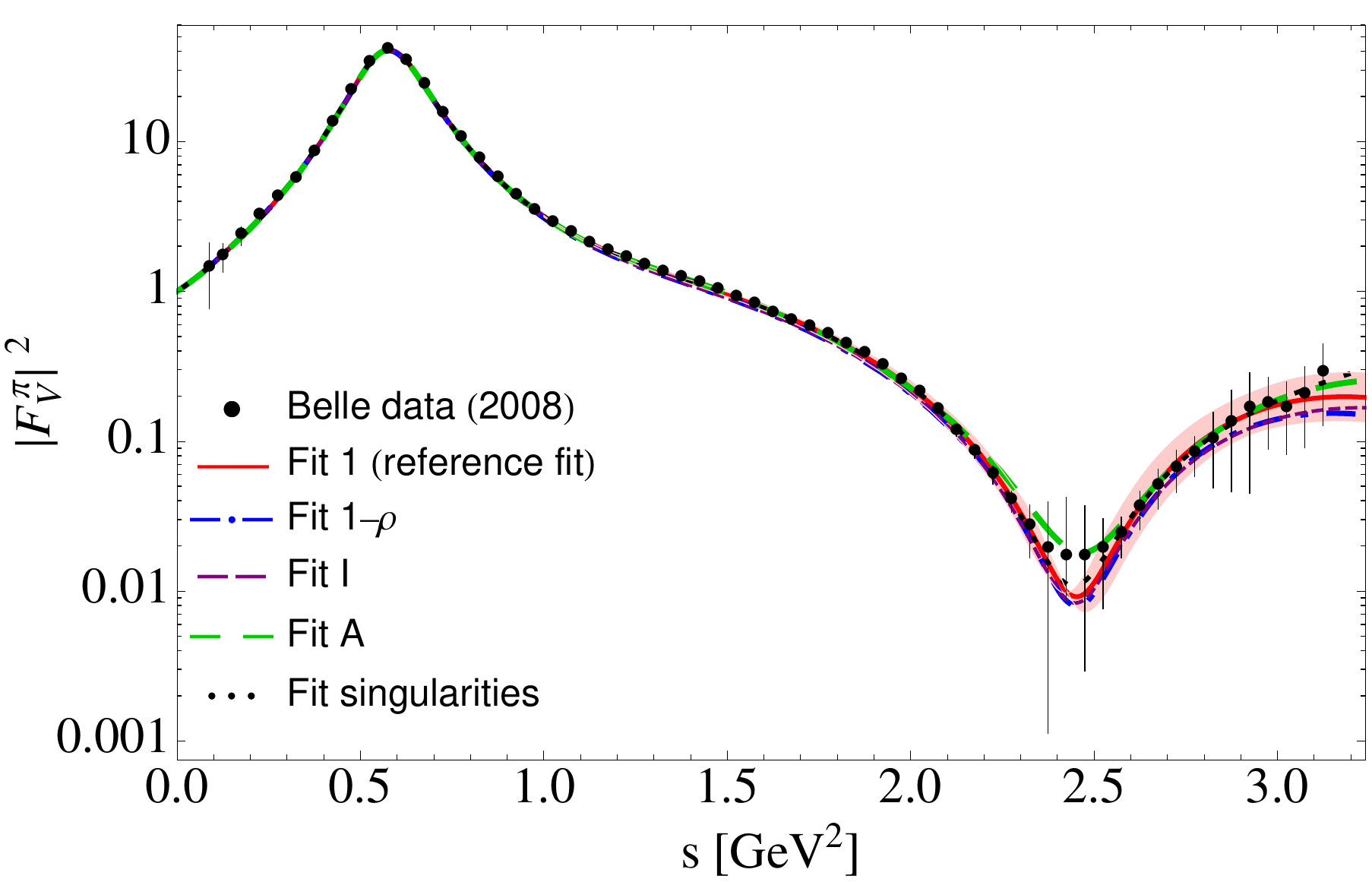}
\caption{\label{FitDispersiveComparisonFF}Belle measurement of the absolute value squared of the pion vector form factor $|F_{V}^{\pi}|^{2}$ (black filled circles) \cite{Fujikawa:2008ma} as compared to our reference fit in Table\,\ref{FitBelle}, Fit 1 (solid red curve), and to our Fit 1-$\rho$ at the matching point of 0.85 GeV (dot-dashed blue curve), Fit I (short dashed purple curve), Fit A (long dashed green curve) and with the fit that avoids singularities (dotted black curve).
See main text for details.}
\end{center}
\end{figure}

We are now in the position to combine all the results from the different fits that we have obtained from our dedicated analysis discussed above and that we graphically compare in Figs.\,\ref{FitDispersiveComparisonPhase} and \ref{FitDispersiveComparisonFF} for the form factor phase shift and modulus squared, respectively.
In particular, we show Fit 1 (reference fit) and Fit 1-$\rho$ with $s_{\rm{cut}}=4$ GeV$^{2}$ from Table \ref{FitBelle}, Fit I at the matching point of 0.85 GeV from Table \ref{FitBelleMatching}, Fit A of Table \ref{CentralFitDispersiveWidths} and the fit results of Eq.\,(\ref{FitBelleCutoff1}) named as Fit singularities.
In the figures, the statistical uncertainty associated to Fit 1 (reference fit) is also displayed by the light red error band.
Tables with the corresponding numerical values including both statistical and systematic errors are given as ancillary material (see Appendix \ref{Supplementary material}).
As seen from these figures, both the phase and the form factor absolute value squared are rather stable and only small differences are seen in the dip region $\sim2.5$ GeV$^{2}$ caused by the destructive interference between the $\rho^{\prime}$ and $\rho^{\prime\prime}$ resonances.
To present our central results, we quote the values of our reference fit in Table\,\ref{FitBelle} with $s_{\rm{cut}}=4$ GeV$^{2}$ (Fit 1) and ascribe a conservative systematic uncertainty coming from the largest variations of central values with respect to the differing results shown in Tables \ref{FitBelle} and \ref{FitBelleMatching} while changing $s_{\rm{cut}}$ and the matching point with the Roy equations, respectively, in Table \ref{CentralFitDispersiveWidths} due to the inclusion of the $K\bar{K}$ decay channel into the $\rho^{\prime}$ and $\rho^{\prime\prime}$ widths and in Eq.\,(\ref{FitBelleCutoff1}) due to the parametrization presented in Eq.\,(\ref{FormFactorCutoffFinal}) that avoids the aforementioned singularities. 
We then obtain
\begin{eqnarray}
& &\alpha_{1}\,=\,1.88\pm0.01\pm0.01\,\,\rm{GeV^{-2}},\quad \alpha_{2}\,=\,4.34\pm0.01\pm0.03\,\,\rm{GeV^{-4}},\nonumber\\[2mm]
& & M_{\rho}\,\doteq\,773.6\pm0.9\pm0.3\,\,\rm{MeV}\,,\nonumber\\[2mm]
& & M_{\rho^{\prime}}\,=\,1376\pm6^{+18}_{-73}\,\,\rm{MeV}\,,\quad\Gamma_{\rho^{\prime}}\,=\,603\pm22_{-141}^{+236}\,\,\rm{MeV}\,,\nonumber\\[2mm]
& & M_{\rho^{\prime\prime}}\,=\,1718\pm4^{+57}_{-94}\,\,\rm{MeV}\,,\quad\Gamma_{\rho^{\prime\prime}}\,=\,465\pm9^{+137}_{-53}\,\,\rm{MeV}\,,\nonumber\\[2mm]
& &\gamma\,=\,0.15\pm0.01^{+0.07}_{-0.03}\,,\quad\phi_{1}\,=\,-0.66\pm0.01^{+0.22}_{-0.99}\,,\nonumber\\[2mm]
& &\delta\,=\,-0.13\pm0.01^{+0.00}_{-0.05}\,,\quad\phi_{2}\,=\,-0.44\pm0.03^{+0.06}_{-0.90}\,,
\label{CentralFitDispersiveScut}
\end{eqnarray}
where the first uncertainty is the statistical fit error while the second is our estimated systematic uncertainty.

As has been already stated in section \ref{exponentialrepresentation}, the resonance mass and width parameters of Eq.\,(\ref{CentralFitDispersiveScut}) are unphysical fit parameters.
To obtain the physical resonance mass and width, we calculate the pole positions in the complex $s$-plane.
This yields:
\begin{eqnarray}
& & M^{\rm{pole}}_{\rho}\,=\,760.6\pm0.8\,\,\rm{MeV}\,,\quad \Gamma^{\rm{pole}}_{\rho}\,=\,142.0\pm0.4\,\,\rm{MeV}\,,\nonumber\\[2mm]
& & M^{\rm{pole}}_{\rho^{\prime}}\,=\,1289\pm8^{+52}_{-143}\,\,\rm{MeV}\,,\quad \Gamma^{\rm{pole}}_{\rho^{\prime}}\,=\,540\pm16^{+151}_{-111}\,\,\rm{MeV}\,,\nonumber\\[2mm]
& & M^{\rm{pole}}_{\rho^{\prime\prime}}\,=\,1673\pm4^{+68}_{-125}\,\,\rm{MeV}\,,\quad \Gamma^{\rm{pole}}_{\rho^{\prime\prime}}\,=\,445\pm8^{+117}_{-49}\,\,\rm{MeV}\,,
\label{Polesfitpipi}
\end{eqnarray}
where the systematic uncertainties are calculated by assuming a Gaussian error propagation while simultaneously varying the corresponding unphysical mass and width given in Eq.\,(\ref{CentralFitDispersiveScut}). 
The results given in Eq.\,(\ref{Polesfitpipi}) constitute one of the fundamental results of the article, we show that the extraction of the pole mass and width of the $\rho^{\prime}$ and $\rho^{\prime\prime}$ resonances is limited by theoretical errors that, as we will see in the following, have been usually ignored or underestimated in the literature.

In Table \ref{ResonancesComparison}, we show our results for the $\rho(1450)$ and $\rho(1700)$ resonance parameters compared to other determinations quoted in the literature.
We consider the pole mass and width as the relevant resonance properties since one expects the pole parameters to be essentially model independent. 
To shed further light on the comparison, we have calculated the pole mass and width of the $\rho(1450)$ and $\rho(1700)$ parameters appearing in phenomenological amplitudes where resonances are introduced through Breit-Wigner (BW) type functions, thus being strongly model dependent.  
This is the case of the Gounaris and Sakurai parametrization \cite{Gounaris:1968mw} used in the fits by ALEPH \cite{Schael:2005am} and Belle \cite{Fujikawa:2008ma}, and also in \cite{Celis:2013xja} where a BW function supplemented by a dispersion relation was used to fit Belle data.
For the $\rho(1450)$, we obtain a pole mass(width) on the lower(upper) side, but in general agreement with respect to previous determinations.
Our results are found to be in a remarkable agreement with \cite{Celis:2013xja}, especially after the pole position is computed, while they are seen in a mild tension with respect to the PDG educated guess reported values $M_{\rho^{\prime}}^{\rm{PDG}}=1450\pm25$ MeV and $\Gamma_{\rho^{\prime}}^{\rm{PDG}}=400\pm60$ MeV \cite{PhysRevD.98.030001}.
For the $\rho(1700)$ we obtain, on the one hand, a pole mass slightly lower than, but in agreement with, previous determinations and the PDG reported value $M_{\rho^{\prime\prime}}^{\rm{PDG}}=1720\pm20$ MeV, and $\sim300$ MeV higher than \cite{Celis:2013xja}.
On the other hand, as seen from the results quoted in the table, the values for the $\rho(1700)$ width show some scatter.
Our value is found to be 250-300 MeV higher than the Belle value and than the findings of \cite{Dumm:2013zh}, $\sim$200 MeV higher than the PDG value $\Gamma_{\rho^{\prime}}^{\rm{PDG}}=250\pm100$ MeV, in line with \cite{Bartos:2017oam} and $\sim$150 MeV lower than \cite{Celis:2013xja}.
Due to the large uncertainties associated to the $\rho(1450)$ and $\rho(1700)$ resonances parameters, however, we cannot draw any more definite conclusion.
For that, more precise data in the $\rho(1450)$ and $\rho(1700)$ region would be highly desirable.

\begin{table}
{\small{ 
\centering
\begin{tabular}{llllll}
\hline
Reference&Model parameters&Pole parameters&Data\\
&$M_{\rho^{\prime}},\Gamma_{\rho^{\prime}}$ [MeV]&$M^{\rm{pole}}_{\rho^{\prime}},\Gamma^{\rm{pole}}_{\rho^{\prime}}$ [MeV]\\
\hline
ALEPH\,\cite{Schael:2005am}&$1328\pm15,468\pm41$&$1268\pm19,429\pm31$&$\tau$\\
ALEPH\,\cite{Schael:2005am}&$1409\pm12,501\pm37$&$1345\pm15,459\pm28$&$\tau\,\&\,e^{+}e^{-}$\\
Belle (fixed $|F_{V}^{\pi}(0)|^{2}$)\,\cite{Fujikawa:2008ma}&$1446\pm7\pm28,434\pm16\pm60$&$1398\pm8\pm31,408\pm13\pm50$&$\tau$\\
Belle (all free)\,\cite{Fujikawa:2008ma}&$1428\pm15\pm26,413\pm12\pm57$&$1384\pm16\pm29,390\pm10\pm48$&$\tau$\\
Dumm et.\,al.\,\cite{Dumm:2013zh}&\quad\quad\quad\quad---&$1440\pm80,320\pm80$&$\tau$\\
Celis et.\,al.\,\cite{Celis:2013xja}&$1497\pm7,785\pm51$&$1278\pm18,525\pm16$&$\tau$\\
Bartos et.\,al.\,\cite{Bartos:2017oam}&\quad\quad\quad\quad---&$1342\pm47,492\pm138$&$e^{+}e^{-}$\\
Bartos et.\,al.\,\cite{Bartos:2017oam}&\quad\quad\quad\quad---&$1374\pm11,341\pm24$&$\tau$\\
{\bf{This work}}&$1376\pm6^{+18}_{-73},603\pm22^{+236}_{-141}$&$1289\pm8^{+52}_{-143},540\pm16^{+151}_{-111}$&$\tau$\\
\hline 
Reference&Model parameters&Pole parameters&Data\\
&$(M_{\rho^{\prime\prime}},\Gamma_{\rho^{\prime\prime}})$ [MeV]&$(M^{\rm{pole}}_{\rho^{\prime\prime}},\Gamma^{\rm{pole}}_{\rho^{\prime\prime}})$ [MeV]\\
\hline
ALEPH\,\cite{Schael:2005am}&$=1713,=235$&$1700,232$&$\tau$\\
ALEPH\,\cite{Schael:2005am}&$1740\pm20,=235$&$1728\pm 20,232$&$\tau\,\&\,e^{+}e^{-}$\\
Belle (fixed $|F_{V}^{\pi}(0)|^{2}$)\,\cite{Fujikawa:2008ma}&$1728\pm17\pm89,164\pm21^{+89}_{-26}$&$1722\pm18,163\pm21^{+88}_{-27}$&$\tau$\\
Belle (all free)\,\cite{Fujikawa:2008ma}&$1694\pm41,135\pm36^{+50}_{-26}$&$1690\pm94,134\pm36^{+49}_{-28}$&$\tau$\\
Dumm et.\,al.\,\cite{Dumm:2013zh}&\quad\quad\quad\quad---&$1720\pm90,180\pm90$&$\tau$\\
Celis et.\,al.\,\cite{Celis:2013xja}&$1685\pm30,800\pm31$&$1494\pm37,600\pm17$&$\tau$\\
Bartos et.\,al.\,\cite{Bartos:2017oam}&\quad\quad\quad\quad---&$1719\pm65,490\pm17$&$e^{+}e^{-}$\\
Bartos et.\,al.\,\cite{Bartos:2017oam}&\quad\quad\quad\quad---&$1767\pm52,415\pm120$&$\tau$\\
{\bf{This work}}&$1718\pm4^{+57}_{-94},465\pm9^{+137}_{-53}$&$1673\pm4^{+68}_{-125},445\pm8^{+117}_{-49}$&$\tau$\\
\hline 
\end{tabular}
\caption{Comparison between different results for the model parameters and corresponding pole positions for the $\rho^{\prime}$ (upper table) and $\rho^{\prime\prime}$ (lower table) resonances.
The first and second errors refer, respectively, to the statistical and systematic uncertainties.}
\label{ResonancesComparison}
}}
\end{table}

Regarding the chiral observables associated to the low-energy expansion of the form factor (cf.\,Eqs.\,(\ref{FFexpansion}) and (\ref{FFexpansion2})), taking into account the results quoted in Eq.\,(\ref{CentralFitDispersiveScut}), we obtain the values shown in the last row of Table \ref{LowEnergyObservables} where we have added the systematic error in quadrature to the statistical uncertainty.
In this table, we also display previous determinations of these quantities for comparison.
\begin{table}
\centering
\begin{tabular}{llllll}
\hline
Reference&$\langle r^{2}\rangle_{V}^{\pi}$ [GeV$^{-2}$]&$c_{V}^{\pi}$ [GeV$^{-4}$]\\
\hline
Colangelo et.\,al.\,\cite{Colangelo:1996hs}&$11.07\pm0.66$&$3.2\pm1.03$\\
Bijnens et.\,al.\,\cite{Bijnens:1998fm}&$11.22\pm0.41$&$3.85\pm0.60$\\
Pich et.\,al.\,\cite{Pich:2001pj}&$11.04\pm0.30$&$3.79\pm0.04$\\
Bijnens et.\,al.\,\cite{Bijnens:2002hp}&$11.61\pm0.33$&$4.49\pm0.28$\\
de Troconiz et.\,al.\,\cite{deTroconiz:2004yzs}&$11.10\pm0.03$&$3.84\pm0.02$\\
Masjuan et.\,al.\,\cite{Masjuan:2008fv}&$11.43\pm0.19$&$3.30\pm0.33$\\
Guo et.\,al.\,\cite{Guo:2008nc}&\quad\quad\,\,---&$4.00\pm0.50$\\
Lattice \cite{Aoki:2009qn}&$10.50\pm1.12$&$3.22\pm0.40$\\
Ananthanarayan et.\,al.\,\cite{Ananthanarayan:2011xt}&$11.17\pm0.53$&$[3.75,3.98]$\\
Ananthanarayan et.\,al.\,\cite{Ananthanarayan:2013dpa}&$[10.79,11.3]$&$[3.79,4.00]$\\
Schneider et.\,al.\,\cite{Schneider:2012ez}&$10.6$&$3.84\pm0.03$\\
Dumm et.\,al.\,\cite{Dumm:2013zh}&$10.86\pm0.14$&$3.84\pm0.03$\\
Celis et.\,al.\,\cite{Celis:2013xja}&$11.30\pm0.07$&$4.11\pm0.09$\\
Ananthanarayan et.\,al.\,\cite{Ananthanarayan:2017efc}&$11.10\pm0.11$&\quad\quad---\\
Hanhart et.\,al.\,\cite{Hanhart:2016pcd}&$11.34\pm0.01\pm0.01$&\quad\quad---\\
Colangelo et.\,al.\,\cite{Colangelo:2018mtw}&$11.02\pm0.10$&\quad\quad---\\
PDG\,\cite{PhysRevD.98.030001}&$11.61\pm0.28$&\quad\quad---\\
{\bf{This work}}&$11.28\pm0.08$&$3.94\pm0.04$\\
\hline 
\end{tabular}
\caption{Low-energy observables of the pion vector form factor up to the quadratic term.
Statistical and systematic uncertainties have been added in quadrature.
Some of the values of the charged pion radius $\langle r^{2}\rangle_{V}^{\pi}$ given in the table are not quoted in the original literature in units of GeV$^{-2}$ but rather in fm$^{2}$ and the conversion has been evaluated by us.}
\label{LowEnergyObservables}
\end{table}
As can be seen, our results are found to be in good agreement with, but in general more precise than, all previous determinations.

It is opportune to mention again that we have treated the subtraction constants $\alpha_{1}$ and $\alpha_{2}$ as free parameters that capture our ignorance of the higher energy part of the integral.
However, in order to check the consistency of our approach, we have also calculated these constants through the sum rule given in Eq.\,(\ref{SumRule}), as for the central values of our analysis presented in Eq.\,(\ref{CentralFitDispersiveScut}), for three different values of $s_{\rm{cut}}$ i.e. 4 GeV$^{2}$, 10 GeV$^{2}$ and $\infty$.
The values we get are collected in Table \ref{SubConstantsSumRule} and are seen in a reasonable good agreement with the results of our fits that we show in the last column for ease of comparison.
In other words, this tell us that the content of the phase is such that saturates rather well the dispersive integral, otherwise the differing results between the sum rules and the fitted subtraction constants would be larger.
\begin{table}[h!]
\begin{center}
  \begin{tabular}{|l|l|l|l|l|c|c|c|c|c|c|c|c|c|c|c|c|c|}
\hline
   \multirow{2}{*}{Sum rule} &  
    \multicolumn{3}{c|}{\multirow{1}{*}{{$s_{\rm{cut}}$ [GeV$^{2}$]}}}& \\ 
    \cline{2-4}
     &4 & 10 & $\infty$& Fit Eq.\,(\ref{CentralFitDispersiveScut})\\ \cline{1-5}
$\alpha_{1}$ &$1.52$&$1.66$&$1.75$&$1.88\pm0.01\pm0.01$\cr
$\alpha_{2}$  &$4.26$&$4.30$&$4.31$&$4.34\pm0.01\pm0.03$\cr
  \hline     
  \end{tabular}
\caption{\label{SubConstantsSumRule}\small{Values for the subtraction constants calculated from the sum rule Eq.\,(\ref{SumRule}), as for the results of our fits given in Eq.\,(\ref{CentralFitDispersiveScut}), for three different values of $s_{\rm{cut}}$ in the dispersive integral.}}
\end{center}
\end{table}

The next shape parameter in the expansion, the cubic slope $d_{V}^{\pi}$, is much less known.
To the best of our knowledge, there are neither theoretical results from ChPT nor calculations on the Lattice.
We obtain 
\begin{eqnarray}
d_{V}^{\pi}&=&10.54\pm0.05\,\,\rm{GeV}^{-6}\,,
\end{eqnarray}
a value which is seen slightly larger than previous estimates $d_{V}^{\pi}=9.70\pm0.40$ GeV$^{-6}$ \cite{Truong:1998yx}, $d_{V}^{\pi}=9.84\pm0.05$ GeV$^{-6}$ \cite{Dumm:2013zh} and $d_{V}^{\pi}=[10.14,10.56]$ GeV$^{-6}$ \cite{Ananthanarayan:2013dpa}.

\section{Predictions and fits to $\tau^{-}\to K^{-}K_{S}\nu_{\tau}$ BaBar data}\label{section3}

The theoretical expression for the differential decay distribution for the transition $\tau^{-}\to K^{-}K^{0}\nu_{\tau}$ in terms of the $K^{-}K^{0}$ invariant mass can be written as \cite{Bruch:2004py,Li:1996md,Palomar:2002hp}

\begin{equation}
\frac{d\Gamma(\tau^{-}\to K^{-}K^{0}\nu_{\tau})}{d\sqrt{s}}=\frac{G_{F}^{2}|V_{ud}|^{2}}{768\pi^{3}}M_{\tau}^{3}\left(1-\frac{s}{M_{\tau}^{2}}\right)^{2}\left(1+\frac{2s}{M_{\tau}^{2}}\right)\sigma_{K}^{3}(s)|F_{V}^{K}(s)|^{2}\,,
\label{DecayDist}
\end{equation}
and it is related to the normalized invariant mass spectrum through
\begin{equation}
\frac{1}{N_{\rm{events}}}\frac{{\rm{d}}N_{\rm{events}}}{{\rm{d}}m_{K^{-}K_{S}}}=\frac{1}{2}\frac{{\rm{d}}\Gamma(\tau^{-}\to K^{-}K_{S}\nu_{\tau})}{{\rm{d}}m_{K^{-}K_{S}}}\frac{1}{\Gamma_{\tau}\bar{B}(\tau^{-}\to K^{-}K_{S}\nu_{\tau})}\Delta^{\rm{bin}}_{K^{-}K_{S}}\,,
\end{equation} 
where $N_{\rm{events}}$ is the total number of measured events, $\Gamma_{\tau}$ is the inverse $\tau$ lifetime and $\Delta^{\rm{bin}}_{K^{-}K_{S}}$ is the bin width.
$\bar{B}(\tau^{-}\to K^{-}K_{S}\nu_{\tau})\equiv\bar{B}$ is a normalization constant that, for a perfect description of the spectrum, would equal the branching ratio. 
For our analysis, we fix this normalization to the BaBar measured branching fraction $\bar{B}=0.739(11)_{\rm{stat}}(20)_{\rm{syst}}\times10^{-3}$ \cite{BaBar:2018qry} \footnote{Another possibility would be to let this constant float and infer its value from fits to the data \cite{Boito:2008fq,Boito:2010me,Escribano:2014joa}.
However, in order to reduce the number of free parameters to fit, we prefer to fix this constant to the branching ratio measured by BaBar.}.
The factor $1/2$ is due to the $K^{-}K_{S}$ decay channel is analyzed.
The corresponding number of events measured by BaBar is $223741\pm3461$ \cite{BaBar:2018qry} and the bin width is $0.04$ GeV.

$F_{V}^{K}(s)$ in Eq.\,(\ref{DecayDist}) denotes the participant kaon vector form factor that we will describe in the following. 
Similar to Eq.\,(\ref{HadronicMatrixElementPion}), it can be defined via the matrix element of the vector current between the vacuum and the $K^{-}K^{0}$ pair as
\begin{equation}
\langle K^{0}K^{-}|\bar{u}\gamma^{\mu}d|0\rangle=\frac{1}{2}\left(p_{K^{0}}-p_{K^{-}}\right)^{\mu}F_{V}^{K}(s)\,,
\end{equation}
where, as in the case of the pion vector form factor, the two-kaon final state corresponds to a $I=J=1$ configuration. 
In order to obtain the expression for $F_{V}^{K}(s)$ at $\mathcal{O}(p^{4})$ in ChPT we need the expressions of the $K^{+}K^{-}$ and $K^{0}\bar{K}^{0}$ form factors.
These can be found in the literature and read \cite{Gasser:1984ux}
\begin{eqnarray}
\label{KKcharged}
F_{K^{+}K^{-}}(s)|_{\rm{ChPT}}&=&1+\frac{2L_{9}^{r}}{F_{\pi}^{2}}-\frac{s}{192\pi^{2}F_{\pi}^{2}}\left[A_{\pi}(s,\mu^{2})+2A_{K}(s,\mu^{2})\right]\,,\\[2mm]
F_{K^{0}\bar{K}^{0}}(s)|_{\rm{ChPT}}&=&-\frac{s}{192\pi^{2}F_{\pi}^{2}}\left[A_{\pi}(s,\mu^{2})-A_{K}(s,\mu^{2})\right]\,.
\label{KKneutral}
\end{eqnarray}
The $I=1$ component corresponding to the $K^{-}K^{0}$ state can be extracted from Eqs.\,(\ref{KKcharged}) and (\ref{KKneutral}) and yields
\begin{equation}
F_{V}^{K}(s)=F_{K^{+}K^{-}}(s)-F_{K^{0}\bar{K}^{0}}(s)=1+\frac{2L_{9}^{r}}{F_{\pi}^{2}}-\frac{s}{96\pi^{2}F_{\pi}^{2}}\left[A_{\pi}(s,\mu^{2})+\frac{1}{2}A_{K}(s,\mu^{2})\right]\,.
\label{FFChPTKaon}
\end{equation} 
We would like to note that the ChPT calculation at $\mathcal{O}(p^{4})$ of the pion (cf.\,Eq.\,(\ref{FFChPT})) and kaon (cf.\,Eq.\,(\ref{FFChPTKaon})) vector form factors are the same.
For our study, we will consider exact $SU(3)$ flavor symmetry and assume that, in a first approximation, both form factors are also the same at energies higher than the chiral region.
This allows us to predict the $\tau^{-}\to K^{-}K_{S}\nu_{\tau}$ decay spectrum from the description of the pion vector form factor carried out in the previous section.

\begin{figure}[thb]
\begin{center}
\vspace*{1.25cm}
\includegraphics[scale=0.85]{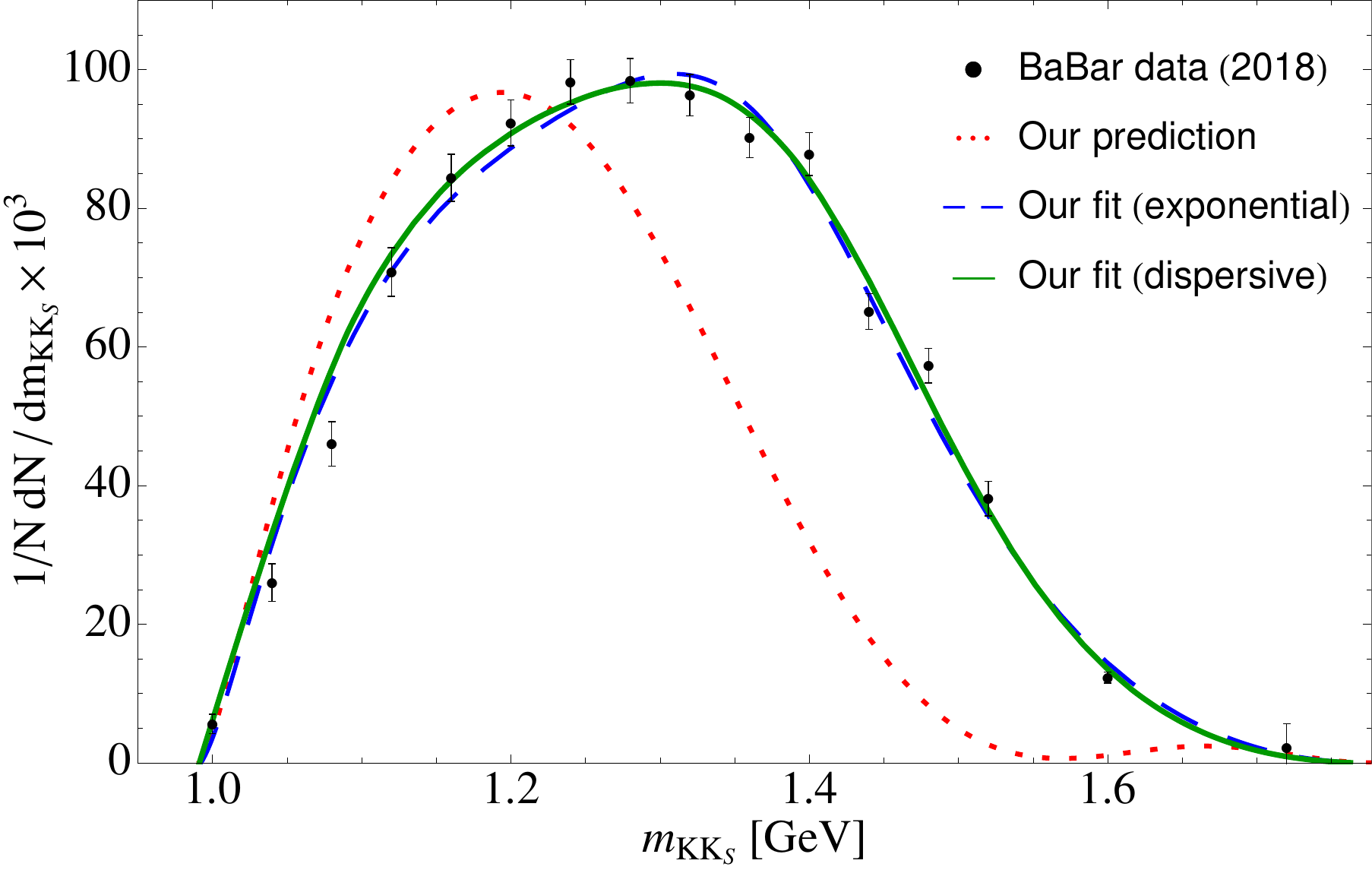}
\caption{\label{SpectrumKKchannel}BaBar data \cite{BaBar:2018qry} for the decay $\tau^-\to K^{-}K_{S}\nu_{\tau}$ (black solid circles) confronted to our prediction (dotted red curve), as for our central results of the pion vector form factor analysis presented in Eq.\,(\ref{CentralFitDispersiveScut}), and fits with the exponential representation (dashed blue curve) and the dispersive approach with $s_{\rm{cut}}=4$ GeV$^{2}$ (solid green curve).}
\end{center}
\end{figure}

In Fig.\,\ref{SpectrumKKchannel}, we show such a prediction (dotted red curve) based on our central results of the pion vector form factor analysis presented in Eq.\,(\ref{CentralFitDispersiveScut}), confronted to the $\tau^{-}\to K^{-}K_{S}\nu_{\tau}$ spectrum measured by BaBar (black circles).
A look at the figure shows a clear disagreement between our prediction and the BaBar data.
The shape of the decay spectrum is not followed by this approach, as also indicated by the low value of the corresponding branching ratio, ${\rm{BR}}(\tau^-\to K^{-}K_{S}\nu_{\tau})=0.545(32)\times10^{-3}$, which is seen $\sim5\sigma$ away the BaBar measurement ${\rm{BR}}(\tau^-\to K^{-}K_{S}\nu_{\tau})=0.739(11)_{\rm{stat}}(20)_{\rm{syst}}\times10^{-3}$.
From these results, we conclude that using the pion vector form factor to describe the $\tau^-\to K^{-}K_{S}\nu_{\tau}$ decay channel is a too rough approach.
This is due to the $K^{-}K^{0}$ production threshold is around $1$ GeV, and thus one should expect the $\rho^{\prime}$ and $\rho^{\prime\prime}$ vector resonances to play a significantly different role than in $\tau^{-}\to\pi^{-}\pi^{0}\nu_{\tau}$ as also noted in \cite{Czyz:2010hj,Volkov:2016yil}.
Thus, the weight of each resonance contribution, represented by the coefficients $\gamma$ and $\delta$ and the phases $\phi_{1}$ and $\phi_{2}$ in Eq.\,(\ref{FFExpThreeRes}), can vary with respect to those entering the form factor of the pion. 
We will check this in the following by performing individual fits to the $\tau^-\to K^{-}K_{S}\nu_{\tau}$ spectrum.
To this end, we adapt the Omn\`{e}s representation of the pion vector form factor (cf. Eq.\,(\ref{FFExpThreeRes})) to the kaon vector form factor ones and write:
\begin{eqnarray}
F_{V}^{K}(s)&=&\frac{M_{\rho}^{2}+s\left(\tilde{\gamma} e^{i\tilde{\phi}_{1}}+\tilde{\delta} e^{i\tilde{\phi}_{2}}\right)}{M_{\rho}^{2}-s-iM_{\rho}\Gamma_{\rho}(s)}\exp\Bigg\lbrace {\rm{Re}}\Bigg[-\frac{s}{96\pi^{2}F_{\pi}^{2}}\left(A_{\pi}(s)+\frac{1}{2}A_{K}(s)\right)\Bigg]\Bigg\rbrace\nonumber\\
&&-\tilde{\gamma}\frac{s\,e^{i\tilde{\phi}_{1}}}{M_{\rho^{\prime}}^{2}-s-iM_{\rho^{\prime}}\Gamma_{\rho^{\prime}}(s)}\exp\Bigg\lbrace-\frac{s\Gamma_{\rho^{\prime}}(M_{\rho^{\prime}}^{2})}{\pi M_{\rho^{\prime}}^{3}\sigma_{\pi}^{3}(M_{\rho^{\prime}}^{2})}{\rm{Re}}A_{\pi}(s)\Bigg\rbrace\nonumber\\
&&-\tilde{\delta}\frac{s\,e^{i\tilde{\phi}_{2}}}{M_{\rho^{\prime\prime}}^{2}-s-iM_{\rho^{\prime\prime}}\Gamma_{\rho^{\prime\prime}}(s)}\exp\Bigg\lbrace-\frac{s\Gamma_{\rho^{\prime\prime}}(M_{\rho^{\prime\prime}}^{2})}{\pi M_{\rho^{\prime\prime}}^{3}\sigma_{\pi}^{3}(M_{\rho^{\prime\prime}}^{2})}{\rm{Re}}A_{\pi}(s)\Bigg\rbrace\,,
\label{FFExpThreeResKaon}
\end{eqnarray}
where the coefficients $\tilde{\gamma}$ and $\tilde{\delta}$ and the phases $\tilde{\phi}_{1}$ and $\tilde{\phi}_{2}$ account, respectively, for the relative importance between the contributions of the different resonances and the corresponding interference in the $K^{-}K^{0}$ system.

From the kaon vector form factor in Eq.\,(\ref{FFExpThreeResKaon}), we extract its phase through
\begin{equation}
\tan\delta_{1}^{KK}(s)=\frac{{\rm{Im}}F^{K}_{V}(s)}{{\rm{Re}}F^{K}_{V}(s)}\,,
\label{KKphase}
\end{equation}
and this is inserted\footnote{The $\delta_{1}^{KK}$ phase as extracted from Eq.\,(\ref{KKphase}) is also matched to the $\pi\pi$ scattering at 1 GeV as explained along the lines of section \ref{section2}.} into a three-times-subtracted dispersive representation of the form factor
\begin{equation}
F_{V}^{K}(s)=\exp\left[\tilde{\alpha}_{1}s+\frac{\tilde{\alpha}_{2}}{2}s^{2}+\frac{s^{3}}{\pi}\int_{4m_{\pi}^{2}}^{s_{\rm{cut}}}ds^{\prime}\frac{\delta_{1}^{KK}(s^{\prime})}{(s^{\prime})^{3}(s^{\prime}-s-i0)}\right]\,,
\label{FFthreesubKK}
\end{equation}
where $\tilde{\alpha}_{1}$ and $\tilde{\alpha}_{2}$ are two subtraction constants corresponding to the slope and curvature of the form factor of the kaon.

We have considered different fits to the measured $m_{K^{-}K_{S}}$ invariant mass distribution\footnote{We would like to notice here that the last two data points of the BaBar paper's Table II \cite{BaBar:2018qry} have been rescaled to match the paper's figure 12.} and found on the one hand that, in full generality, the data is not sensitive either to the low-energy region or to the $\rho(770)$ peak region.
This is expected due to the $K^{-}K_{S}$ production threshold opens around 1000 MeV which is some 100 MeV larger than $M_{\rho}+\Gamma_{\rho}$, the energy region more influenced by the presence of the $\rho(770)$.
This implies first that the slopes of the kaon vector form factor, which encode the physics immediately above threshold, cannot be fitted with $\tau^-\to K^{-}K_{S}\nu_{\tau}$ data and second that the fits lead unrealistic parameters under floating the $\rho$-meson parameters.
On the other hand, the data is scarce in the $\rho(1700)$ resonance region and thus not suitable for extracting the corresponding resonance parameters.
We have therefore fixed the slopes associated to the kaon vector form factor and the $\rho$-meson mass ($773.6(9)$ MeV), and considered fits varying only the $\rho(1450)$-resonance mass and width, and $\tilde{\gamma}$ and $\tilde{\phi}_{1}$, while neglecting the contribution of the $\rho(1700)$ resonance to the decay i.e. $\tilde{\delta}_{1}=0$.
In Table \ref{DispersiveKK}, we show the results of our fits using different settings.
Fit $i)$ corresponds to fixing the slopes $\tilde{\alpha}_{1}$ and $\tilde{\alpha}_{2}$, to the slopes $\alpha_{1}$ and $\alpha_{2}$ given in Eq.\,(\ref{CentralFitDispersiveScut}) obtained from the analysis of the pion vector form factor, while Fits $ii)$ and $iii)$ are variants of it.
In particular, Fit $ii)$ is the result of fixing the slope to $\tilde{\alpha}_{1}=1.84(30)$ obtained from the kaon radius of Ref.\,\cite{Bijnens:2002hp} while Fit $iii)$ includes the intermediate $K\bar{K}$ state into the $\rho^{\prime}$ decay width (cf.\,Eq.\,(\ref{KKintowidths})). 
Finally, Fit $iv)$ is the result of the direct application of the exponential vector form factor in Eq.\,(\ref{FFExpThreeResKaon}) to fit experimental data.

\begin{table}[h!]
\begin{center}
  \begin{tabular}{|l|l|l|l|l|l|c|c|c|c|c|c|c|c|c|c|c|c|c|}
\hline
    \multirow{2}{*}{Parameter} &  
    \multicolumn{4}{c|}{\multirow{1}{*}{{$s_{\rm{cut}}=4$ [GeV$^{2}$]}}} \\ 
\cline{2-5}
     &Fit $i)$ & Fit $ii)$ &Fit $iii)$&Fit $iv)$\\ \cline{1-5}
$\tilde{\alpha}_{1}$  &$=1.88(1)$&$=1.84$&$=1.88(1)$&---\cr
$\tilde{\alpha}_{2}$  &$=4.34(1)$&$=4.34$&$=4.34(1)$&---\cr
$M_{\rho^{\prime}}$ [MeV]&$1467(24)$&$1538(32)$&$1489(25)$&1411(12)\cr
$\Gamma_{\rho^{\prime}}$ [MeV]&$415(48)$&$604(83)$&$297(36)$&394(35)\cr
$\tilde{\gamma}$ &$0.10(2)$&$0.36(11)$&$0.10(2)$&0.09(1)\cr
$\tilde{\phi}_{1}$ &$-1.19(16)$&$-1.48(13)$&$-1.10(15)$&$-1.88(9)$\cr
\hline
$\chi^{2}$/d.o.f. &$2.9$&$1.9$&$2.9$&3.3\cr
\cline{1-5}
  \end{tabular}
\caption{\label{DispersiveKK}\small{Results for the fit to the BaBar $\tau^{-}\to K^{-}K_{S}\nu_{\tau}$ data \cite{BaBar:2018qry} with a three-times-subtracted dispersion relation including two vector resonances in $F_{V}^{K}(s)$ according to Eq.\,(\ref{FFthreesubKK}) with $s_{\rm{cut}}=4$ GeV$^{2}$ in the dispersive integral.}}
\end{center}
\end{table}


The fit with best $\chi^{2}$/d.o.f is seen for Fit $ii)$ but the resulting fit parameters carry the larger error, and the (large) uncertainty associated to $\tilde{\alpha}_{1}$ has not been taken into account.
Because of that, the results from Fit $ii)$ should be taken with a word of caution and we only consider them as illustration of the potential effects due to the low-energy parameters of the kaon form factor. 
The values appearing in Table \ref{DispersiveKK} can be translated to pole values along the lines discussed in the previous section.
This yields $M^{\rm{pole}}_{\rho^{\prime}}\,=\,1422\pm22$ MeV and $\Gamma^{\rm{pole}}_{\rho^{\prime}}\,=\,393\pm41$ MeV (Fit $i)$), $M^{\rm{pole}}_{\rho^{\prime}}\,=\,1453\pm29$ MeV and $\Gamma^{\rm{pole}}_{\rho^{\prime}}\,=\,546\pm70$ MeV (Fit $ii)$), and $M^{\rm{pole}}_{\rho^{\prime}}\,=\,1466\pm23$ MeV and $\Gamma^{\rm{pole}}_{\rho^{\prime}}\,=\,289\pm33$ MeV (Fit $iii)$) for the dispersive approaches, and $M^{\rm{pole}}_{\rho^{\prime}}\,=\,1370\pm15$ MeV and $\Gamma^{\rm{pole}}_{\rho^{\prime}}\,=\,373\pm30$ MeV for the exponential representation (Fit $iv)$).
From these results we conclude that while the pole mass of the $\rho(1450)$ resonance as extracted from the $\tau^{-}\to K^{-}K_{S}\nu_{\tau}$ decay tends to be larger than the values obtained in the previous section from the analysis of the pion vector form factor, the width tends to be smaller and the associated fit uncertainties are in both cases larger.
Also, we would like to note that the relative weight $\tilde{\gamma}$ as extracted from the $K^{-}K_{S}$ channel is found to be in accordance with the values determined in the previous section.  

As a matter of example, in Fig.\,\ref{SpectrumKKchannel} we provide a graphical account of the dispersive Fit $i)$ (solid green curve) and of the exponential Fit $iv)$ (dashed blue curve).
Notice that, as occurs in \cite{Luchinsky:2018lfj}, the second and third data points are difficult to accommodate in any case.
These results show that, although the fits to the $\tau^-\to K^{-}K_{S}\nu_{\tau}$ decay spectrum have considerably improved with respect to predictions discussed above (dotted red curve) and seem to agree rather well with BaBar data, yielding BR$(\tau^{-}\to K^{-}K_{S}\nu_{\tau})=0.749(93)\times10^{-3}$ (Fit $i)$) and BR$(\tau^{-}\to K^{-}K_{S}\nu_{\tau})=0.744(89)\times10^{-3}$ (Fit $iv)$), the quality of the fit as indicated by the $\chi^{2}$/d.o.f is not satisfactory enough.
This fact motivates the combined analysis, detailed in the next section, of the Belle data of the pion vector form factor modulus squared $|F_{V}^{\pi}|^{2}$ and the BaBar data of the decay $\tau^{-}\to K^{-}K_{S}\nu_{\tau}$.
Such analysis shall allow us to determine the $\rho(1450)$ and $\rho(1700)$ resonance parameters with improved precision and obtain a good description of the measured $K^{-}K_{S}$ decay spectrum.

\newpage

\section{Joint fits to $|F_{V}^{\pi}|^{2}$ Belle and $\tau^{-}\to K^{-}K_{S}\nu_{\tau}$ BaBar data}\label{section4}

The $\chi^{2}$ minimised in our simultaneous fit is
\begin{equation}
\chi^{2}=\sum_{i}^{62}\left(\frac{|F_{V}^{\pi}(s_{i})|^{2}_{\rm{th}}-|F_{V}^{\pi}(s_{i})|^{2}_{\rm{exp}}}{\sigma_{|F_{V}^{\pi}(s_{i})|^{2}_{\rm{exp}}}}\right)^{2}+\sum_{j}^{16}\left(\frac{\mathcal{N}_{j}^{\rm{th}}-\mathcal{N}_{j}^{\rm{exp}}}{\sigma_{\mathcal{N}_{j}^{\rm{exp}}}}\right)^{2}\,,
\end{equation}
where the first and second terms correspond, respectively, to the Belle pion vector form factor data \cite{Fujikawa:2008ma} and to the BaBar $\tau^{-}\to K^{-}K_{S}\nu_{\tau}$ measurement \cite{BaBar:2018qry}. 
For the later, $\mathcal{N}_{j}^{\rm{exp}}$ and $\sigma_{\mathcal{N}_{j}^{\rm{exp}}}$ are the experimental normalized number of events for $\tau^{-}\to K^{-}K_{S}\nu_{\tau}$ and the associated uncertainties in the $i$-th bin, respectively. 

The parameters entering the dispersive representation of the form factors of Eqs.\,(\ref{FFthreesub}) and (\ref{FFthreesubKK}) are therefore determined by a simultaneous fit to both data sets and include:
\begin{itemize}
\item The two subtraction constants, $\alpha_{1,2}$ and $\tilde{\alpha}_{1,2}$, corresponding to the slope and curvature parameters associated to the low-energy expansion of the pion and kaon form factors.
\item The masses and decay widths of the participating $\rho^{\prime}$ and $\rho^{\prime\prime}$ resonances, $M_{\rho^{\prime},\rho^{\prime\prime}}$ and $\Gamma_{\rho^{\prime},\rho^{\prime\prime}}$, used to model the phase entering the dispersive integral.
The parameter for the $\rho$-meson mass, $M_{\rho}$, is taken equal to that entering the phase shift, $m_{\rho}$, and its quoted value is fixed to $773.6(9)$ MeV as discussed in section \ref{section2}. 
\item The resonance mixing parameters, $\gamma,\delta$ and $\tilde{\gamma},\tilde{\delta}$, and their phases, $\phi_{1,2}$ and $\tilde{\phi}_{1,2}$.
\end{itemize}

In Table \ref{CombinedFit1}, we show the results of our simultaneous fits using slightly different settings, though in all of them a three-times-subtracted dispersion relation according to Eqs.\,(\ref{FFthreesub}) and (\ref{FFthreesubKK}) with $s_{\rm{cut}}=4$ GeV$^{2}$ in the dispersive integral is employed.
Fit $a$ (second column) corresponds to fixing $\tilde{\alpha_{1}}=\alpha_{1}$ and $\tilde{\alpha_{2}}=\alpha_{2}$ and taking $\tilde{\delta}=0$ because, as we have discussed in section \ref{section3}, the BaBar measurement of the decay $\tau^{-}\to K^{-}K_{S}\nu_{\tau}$ is still not sensitive to $\rho(1700)$ resonance properties.
The corresponding fit results supports the relative weights $\gamma$ and $\tilde{\gamma}$ to be the same for the $\pi^{-}\pi^{0}$ and $K^{-}K_{S}$ channels.
This feature is proven in Fit $b$ (third column) by enforcing $\gamma=\tilde{\gamma}$.
By doing this, the $\chi^{2}$/d.o.f is reduced from $1.52$ to $1.19$ and the values of the fitted parameters remain basically the same but for the $\rho^{\prime\prime}$-width, whose central value is shifted downwards by $\sim 100$ MeV, and to less extent for the $\rho^{\prime}$-mass, which suffers a variation of $\sim 50$ MeV upwards, but still compatible within errors.
Finally, Fit $c$ (last column) is the result of letting all parameters to float independently and the corresponding fit parameters are found to be compatible with Fits $a$ and $b$, though with larger uncertainties.
This fit also yields results that supports the assumption $\tilde{\alpha_{1}}=\alpha_{1}$ and $\tilde{\alpha_{2}}=\alpha_{2}$ made in Fits $a$ and $b$.
As a side result, we extract the charge kaon radius through $\langle r^{2}\rangle_{V}^{K}=6\tilde{\alpha}_{1}$ (cf.\,Eq.\,(\ref{FFexpansion2})).
Our value, $\langle r^{2}\rangle_{V}^{K}=(11.28\pm1.44)$ GeV$^{-2}$, lies in the ballpark of results for this quantity $\langle r^{2}\rangle_{V}^{K}=(9.09\pm1.82)$ GeV$^{-2}$ and $\langle r^{2}\rangle_{V}^{K}=(11.07\pm1.82)$ GeV$^{-2}$ \cite{Bijnens:2002hp}, $\langle r^{2}\rangle_{V}^{K}=(9.76\pm0.85)$ GeV$^{-2}$ \cite{Aoki:2015pba} and $\langle r^{2}\rangle_{V}^{K}=[10.02,10.79]$ GeV$^{-2}$ \cite{Krutov:2016luz}.

\begin{table}[h!]
\begin{center}
  \begin{tabular}{|l|l|l|l|l|l|c|c|c|c|c|c|c|c|c|c|c|c|c|}
\hline
    \multirow{2}{*}{Parameter} &  
    \multicolumn{3}{c|}{\multirow{1}{*}{{$s_{\rm{cut}}=4$ [GeV$^{2}$]}}} \\ 
\cline{2-4}
     &Fit $a$ & Fit $b$ &Fit $c$\\ \cline{1-4}
$\alpha_{1}$ &$1.88(1)$&$1.89(1)$&$1.87(1)$\cr
$\alpha_{2}$  &$4.34(2)$&$4.31(2)$&$4.38(3)$\cr
$\tilde{\alpha}_{1}$  &$=\alpha_{1}$&$=\alpha_{1}$&$1.88(24)$\cr
$\tilde{\alpha}_{2}$  &$=\alpha_{2}$&$=\alpha_{2}$&$4.38(29)$\cr
 $m_{\rho}$ [MeV]&$=773.6(9)$&$=773.6(9)$&$=773.6(9)$\cr
 $M_{\rho}$ [MeV]&$=m_{\rho}$&$=m_{\rho}$&$=m_{\rho}$\cr
$M_{\rho^{\prime}}$ [MeV]&$1396(19)$&$1453(19)$&$1406(61)$\cr
$\Gamma_{\rho^{\prime}}$ [MeV]&$507(31)$&$499(51)$&$524(149)$\cr
$M_{\rho^{\prime\prime}}$[MeV] &$1724(41)$&$1712(32)$&$1746(1)$\cr
$\Gamma_{\rho^{\prime\prime}}$ [MeV]&$399(126)$&$284(72)$&$413(362)$\cr
$\gamma$ &$0.12(3)$&$0.15(3)$&$0.11(11)$\cr
$\tilde{\gamma}$ &$0.11(2)$&$=\gamma$&$0.11(5)$\cr
$\phi_{1}$ &$-0.23(26)$&$0.29(21)$&$-0.27(42)$\cr
$\tilde{\phi}_{1}$ &$-1.83(14)$&$-1.48(13)$&$-1.90(67)$\cr
$\delta$ &$-0.09(2)$&$-0.07(2)$&$-0.10(5)$\cr
$\tilde{\delta}$ &$=0$&$=0$&$-0.01(4)$\cr
$\phi_{2}$ &$-0.20(31)$&$0.27(29)$&$-1.15(71)$\cr
$\tilde{\phi}_{2}$ &$=0$&$=0$&$0.40(3)$\cr
\cline{1-4}
$\chi^{2}$/d.o.f&$1.52$&$1.19$&$1.25$\cr
\cline{1-4}
  \end{tabular}
\caption{\label{CombinedFit1}\small{Simultaneous fit results for different choices regarding resonance mixing and linear slopes parameters obtained with a three-times-subtracted dispersion relation including three vector resonances in $F_{V}^{\pi}(s)$ and $F_{V}^{K}(s)$ according to Eqs.\,(\ref{FFthreesub}) and (\ref{FFthreesubKK}) with $s_{\rm{cut}}=4$ GeV$^{2}$ in the dispersive integral.}}
\end{center}
\end{table}

\begin{figure}
\begin{center}
\includegraphics[scale=0.745]{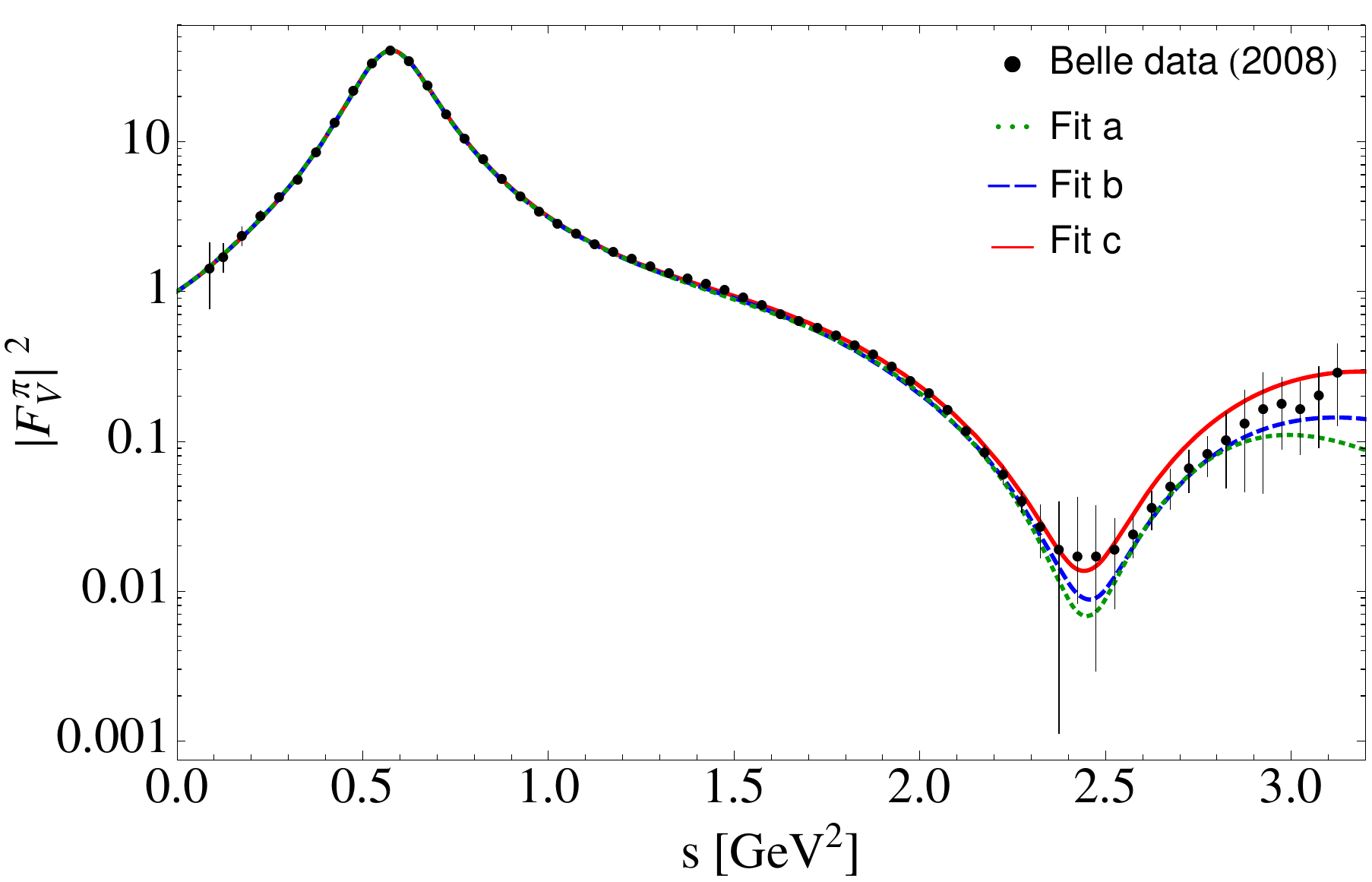}
\caption{\label{JointFit1}Belle measurement of $|F_{V}^{\pi}|^{2}$ (black filled circles) \cite{Fujikawa:2008ma} as compared to our fits obtained from a simultaneous analysis of $|F_{V}^{\pi}|^{2}$ and $\tau^-\to K^{-}K_{S}\nu_{\tau}$ (see Table \ref{CombinedFit1}).}
\end{center}
\end{figure}

\begin{figure}
\begin{center}
\includegraphics[scale=0.745]{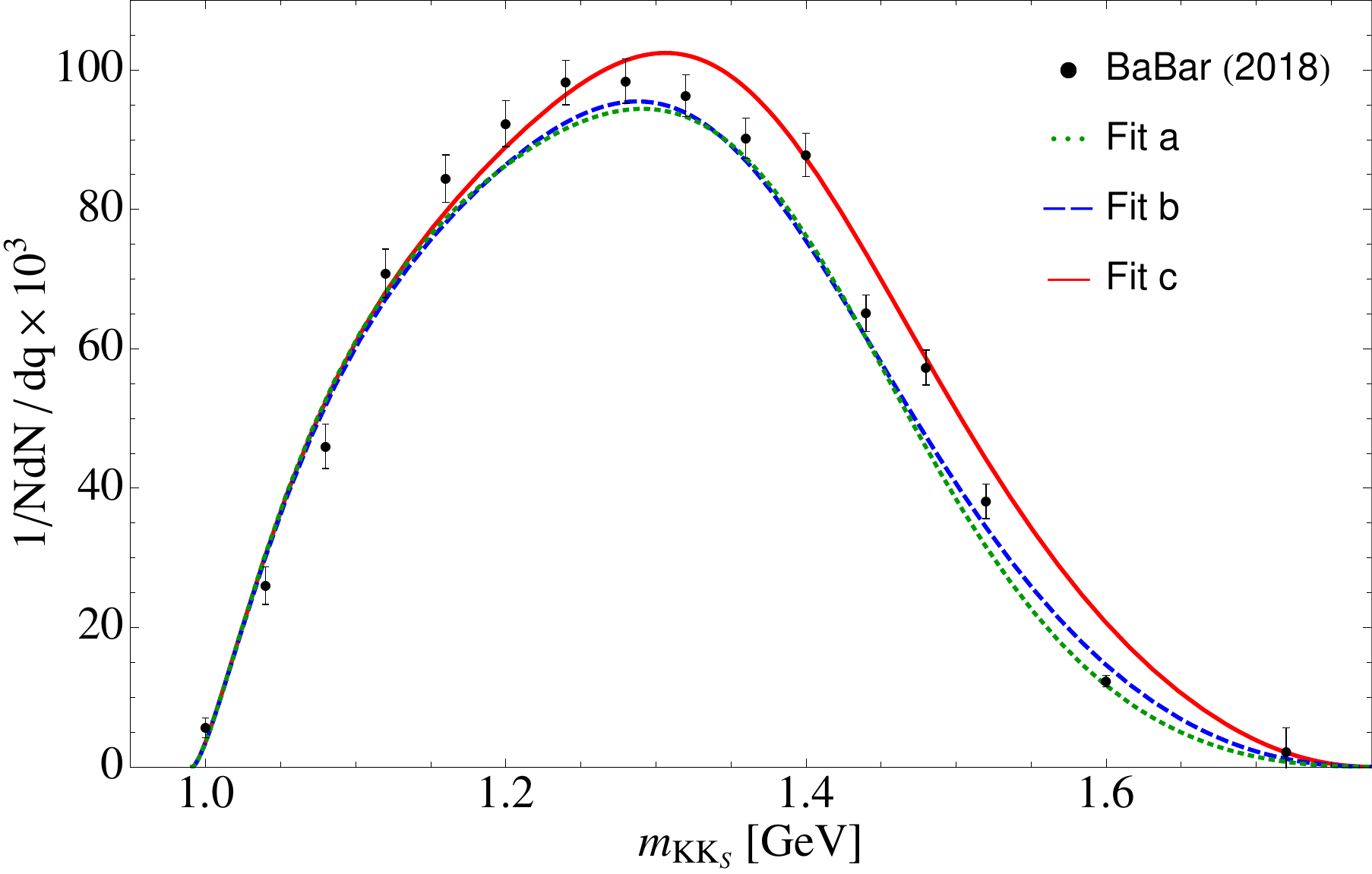}
\caption{\label{JointFit2}BaBar data \cite{BaBar:2018qry} for $\tau^-\to K^{-}K_{S}\nu_{\tau}$ (black solid circles) as compared to our fits obtained from a simultaneous analysis of $|F_{V}^{\pi}|^{2}$ and $\tau^-\to K^{-}K_{S}\nu_{\tau}$ (see Table \ref{CombinedFit1}).}
\end{center}
\end{figure}

The results of our fits are confronted to the Belle $|F_{V}^{\pi}|^{2}$ form factor measurement and to the BaBar $\tau^{-}\to K^{-}K_{S}\nu_{\tau}$ distribution in Figs.\,\ref{JointFit1} and \ref{JointFit2}, respectively.
Satisfactory agreement with the experimental data is seen in accord with the observed $\chi^{2}$/d.o.f.
However, the potential of a combined analysis of the $|F_{V}^{\pi}|^{2}$ and $\tau^-\to K^{-}K_{S}\nu_{\tau}$ data is presently limited by the fact that the errors associated to the latter are still relatively large and the BaBar measurement of the $\tau^-\to K^{-}K_{S}\nu_{\tau}$ spectrum is yet not sensitive to the $\rho(1700)$ resonance properties.
This presents a limitation in determining the $\rho(1450)$ and $\rho(1700)$ resonance parameters with improved precision with respect to the individual analysis of $|F_{V}^{\pi}|^{2}$.
Because of that, we postpone a dedicated study of the systematic uncertainties as we did in section \ref{section2} for the future, when new and more precise measurements become available.

\section{Conclusions}\label{conclusions}

An ideal roadmap for describing meson form factors would require a model-independent approach demanding a full knowledge of QCD in both its perturbative and non-perturbative regimes, knowledge not yet unraveled. 
An alternative to such enterprise would pursuit a synergy between the formal theoretical calculations and experimental data.
In this respect, dispersion relations are a powerful tool to direct oneself towards a model-independent description of form factors.
In this paper, we have revisited the pion vector form factor as extracted from $\tau\to\pi^{-}\pi^{0}\nu_{\tau}$, and provided a parametrization for the kaon vector form factor that describes the decay $\tau^{-}\to K^{-}K_{S}\nu_{\tau}$, by exploiting the synergy between dispersion relations and Chiral Perturbation Theory.

The pion vector form factor is a classic object in low-energy QCD that provides a privileged benchmark to study the effects of $\pi\pi$ interaction under rather clean conditions.
These pion-pion interactions are universal and enter the description of many physical observables, hence the importance of having good control of them.
For our analysis, we have used a three-times-subtractred dispersion relation and exploited Watson's theorem, and the fact that the elastic $P$-wave $\pi\pi$ interactions capturing the physics of the $\rho$-resonance are encoded in the phase shift, to drive the form factor phase entering the dispersive integral up to 1 GeV from the well-known parametrization of the $\pi\pi$ scattering phase shift existent in the literature.
Above 1 GeV, the $\rho(1450)$ and $\rho(1700)$ resonance effects show up and to obtain an improved description of the energy region where these resonances come up into play is one of the purposes of this work.
For that, we have used a model for the phase extracted from the exponential Omn\`{e}s representation of the form factor (cf.\,Eq.\,(\ref{FFExpThreeRes})), whose direct application to the pion vector form factor experimental data is seen very satisfactory (see Fig.\,\ref{FitExp}), that we match smoothly at 1 GeV to the $\pi\pi$ scattering phase.
Armed with this parametrization, we have carried out a very dedicated analysis of the high-statistics Belle experimental data and assessed the role of the theoretical systematic uncertainties in the determination of the $\rho(1450)$ and $\rho(1700)$ physical resonance parameters by considering a number of variants of it. 
Tables with the corresponding numerical values including both statistical and systematic errors are given as ancillary material of this paper (see Appendix \ref{Supplementary material}).
From our study, we conclude that the determination of the pole mass and width of these resonances (cf.\,Eq.\,(\ref{Polesfitpipi})) is limited by theoretical errors that have been usually ignored or underestimated in the literature so far.

On a second stage, we have performed a first analysis of the recent BaBar measurement experimental data on $\tau^{-}\to K^{-}K_{S}\nu_{\tau}$ based on a parametrization of the participant kaon vector form factor that is built in a similar fashion to that of the pion.
We have shown that while the production threshold of this decay channel sits around 1000 MeV and therefore it is out of the $\rho(770)$-dominance region, the role of the $\rho(1450)$-resonance is different than in the $\pi^{-}\pi^{0}$ mode and indeed dominates.
As a result of our fits (see Table \ref{DispersiveKK} and Fig.\,\ref{SpectrumKKchannel}), we have extracted its associated pole parameters.
Regarding the $\rho(1700)$, the data is scarce in this region and thus it is not-yet suitable for extracting the corresponding resonance parameters.

  
Finally, we have pointed out that high-quality data on the decay $\tau^{-}\to K^{-}K_{S}\nu_{\tau}$ will allow one to determine the $\rho(1450)$ and $\rho(1700)$ resonance pole parameters with improved precision from a combined analysis with the pion vector form factor.
In summary, we hope our analysis to be of interest for present and future $Z$, tau-charm and $B$-factories where new measurements should be possible. 

\section*{Acknowledgements}

The authors acknowledge fruitful discussions with Feng-Kun Guo and Bing-Son Zou.
The work of S.G-S has been supported in part by the CAS President's International Fellowship Initiative for Young International Scientists (Grant No.\,2018DM0034), by the Sino-German Collaborative Research Center \textquotedblleft Symmetries and the Emergence of Structure in QCD\textquotedblright\,(NSFC Grant No.\,11621131001, DFG Grant No.\,TRR110), by NSFC (Grant No.\,11747601) and by the National Science Foundation (PHY-1714253). 
P. R. acknowledges support  from  Conacyt  through  projects  FOINS-296-2016  (Fronteras  de  la  Ciencia), 250628 (Ciencia B\'{a}sica) and Fondo SEP-Cinvestav 2018.

\appendix

\section{Supplementary material}\label{Supplementary material}

As supplementary material of this paper, we provide our fit results shown in Figs.\,\ref{FitDispersiveComparisonPhase} and \ref{FitDispersiveComparisonFF} for the pion form factor phase and modulus squared in two data files named FormFactorPhase.dat and FormFactorModulusSquared.dat, respectively.

In the FormFactorPhase.dat file, the first column is $\sqrt{s}$ (in GeV) and the other two columns correspond to the numerical value of the phase and the statistical error for each fit, and they are given until 2 GeV.
For our reference fit (Fit 1), moreover, we provide a fourth column with our conservative estimate for the systematic error.
We calculate this after symmetrizing the asymmetric systematic uncertainties associated to the fit parameters presented in Eq.\,(\ref{CentralFitDispersiveScut}) and by assuming a Gaussian error propagation while simultaneously varying the corresponding parameters.

In the FormFactorModulusSquared.dat file, the first column is $s$ (in GeV$^{2}$) and the second column is the value for the form factor modulus squared for each fit, and is given until $s=m_{\tau}^{2}$.
The associated upper and lower statistical error bands are collected in the third and fourth columns of the file, respectively. 
As before, for our reference fit (Fit 1), we also provide an estimated systematic error, and the resulting upper and lower bands are gathered, respectively, in columns fifth and sixth of the file.

\end{document}